\documentclass[twocolumn,prd,aps,nofootinbib,showpacs,superscriptaddress]{revtex4-1}
\usepackage{longtable}
\usepackage{multirow}
\usepackage{graphicx}
\usepackage[dvips]{color}
\usepackage{amssymb,amsmath}
\usepackage{supertabular}
\allowdisplaybreaks[4]
\begin{document}

\title{Studying the $\rho$ resonance parameters with staggered fermions}

\author{Ziwen Fu}
\email{fuziwen@scu.edu.cn}
\affiliation{
Key Laboratory for Radiation Physics and Technology of  Education Ministry;
Institute of Nuclear Science and Technology, Chengdu 610064, China
}

\affiliation{
Center for Theoretical Physics, College of Physical Science and Technology, Sichuan University,
Chengdu 610064, China
}

\author{Lingyun Wang}
\email{ylingyunwang@yahoo.com}
\affiliation{
International Affair Department, Chengdu Jiaxiang Foreign Language School, Chengdu 610023, China
}

\begin{abstract}
We deliver a lattice study of $\rho$ resonance parameters with $p$-wave $\pi\pi$ scattering phases,
which are extracted by finite-size methods at one center-of-mass frame and
four moving frames for six MILC lattice ensembles with pion masses ranging
from $346$ to $ 176$~MeV.
The effective range formula is applied to describe the scattering phases
as a function of the energy covering the resonance region, this allows us
to extract $\rho$ resonance parameters and to investigate the quark-mass dependence.
Lattice studies with three flavors of the Asqtad-improved staggered fermions
enable us to use the moving-wall source technique on large lattice spatial
dimensions ($L=64$) and small light $u/d$ quarks.
Numerical computations are carried out at two lattice spacings, $a \approx 0.12$ and $0.09$~fm.
\end{abstract}


\maketitle

\section{Introduction}
\label{Sec:Introduction}
Resonances decay into elementary particles via strong interaction,
which is experimentally studied by scattering approaches.
The theoretical computation of the resonance parameters from QCD is difficult because of its nonperturbative property.
At present, it is practical to apply lattice QCD
to calculate the scattering observables.
The $\rho$ meson is the simplest resonance for such a lattice study.
The principal decay channel of the $\rho$ meson
is to a pair of pions with a branching rate close to $100.0\%$~\cite{Agashe:2014kda},
which can be precisely handled with lattice QCD.
Nonetheless, the reliable calculations of $\pi\pi$ correlators are expensive;
hence, the hadronic coupling constants were used
in early studies of the $\rho$  resonance parameters~\cite{Vasanti:1976zz,Gottlieb:1985rc,Loft:1988sy,
Altmeyer:1995qx,McNeile:2002fh,Jansen:2009hr}.

With the great progress of numerical algorithms,
aided by the tremendous advancement of computer power,
the finite-size formula established by L\"uscher in the center-of-mass
frame (CMF)~\cite{Luscher:1990ux,Luscher:1990ck}
and the extensions to the moving frame (MF)~\cite{Rummukainen:1995vs}
have been employed to extract $\rho$ resonance parameters
from $p$-wave $I=1$ $\pi\pi$ scattering phases.
Such an exploratory study was conducted by the CP-PACS Collaboration
with Wilson fermions~\cite{Aoki:2007rd}.
After this pioneering work, the QCDSF Collaboration delivered
results with clover fermions~\cite{Gockeler:2008kc},
the ETMC Collaboration reported results with maximally twisted mass fermions and
explored the pion mass dependence~\cite{Feng:2010es},
J.~Frison {\it et al.} presented preliminary results with Wilson fermions and
pion masses as low as $200$~MeV~\cite{Frison:2010ws},
Lang {\it et al.} delivered results with clover-Wilson fermions using Laplacian Heaviside
smearing operators~\cite{Lang:2011mn},
and the PACS-CS Collaboration investigated them
with Wilson fermions using the efficient smearing techniques~\cite{Aoki:2011yj}.
Pelissier and Alexandru presented results from asymmetrical lattices using nHYP-smeared clover fermions~\cite{Pelissier:2012pi}.
The Hadron Spectrum Collaboration (HSC) adopted the anisotropic
lattice formulation of clover fermions~\cite{Dudek:2012xn}
and recently further used the coupled channel~\cite{Wilson:2015dqa}
to study $\rho$ resonance parameters.
Good statistical precision was obtained from anisotropic Wilson clover
by J. Bulava {\it et al.}~\cite{Bulava:2015qjz}, and
Guo {\it et al.} presented their results with nHYP-smeared clover fermions~\cite{Guo:2016zos}.
The RQCD Collaboration recently computed $\rho$ resonance parameters at a nearly
physical pion mass using nonperturbatively improved Wilson fermions~\cite{Bali:2015gji}.

It is well known that the rectangular diagrams of $I=1$ $\pi\pi$ scattering
are hard to calculate, and the stochastic source method, or its variants (the distillation method, etc.)
are normally used to study the $\rho$ resonance~\cite{Aoki:2007rd,Aoki:2011yj,Gockeler:2008kc,Feng:2010es,Frison:2010ws,Pelissier:2012pi,Dudek:2012xn,Lang:2011mn,Bulava:2015qjz,Guo:2016zos,Wilson:2015dqa,Bali:2015gji}.
Although it is luxury, the moving wall source technique,
which has been explored in the center-of-mass frame~\cite{Kuramashi:1993ka,Fukugita:1994ve},
is believed to calculate the rectangular diagram of
the two-particle scattering with high quality.
Recently we further extended this method to a two-particle system
with nonzero momenta to tentatively investigate
the $\kappa$, $\sigma$,  and $K^\star(892)$ meson decays~\cite{Fu:2012gf},
along with a few studies of the meson-meson scattering~\cite{Fu:2011bz}.
In these works, we found that the moving-wall source can calculate
both the four-point and three-point correlators
with high quality; this encourages us to exploratively use this technique
to study $\rho$ resonance parameters.

The rapid deterioration of the pion propagator signal
as momentum increases is impressive~\cite{DellaMorte:2012xc}.
According to the analytical arguments in Ref.~\cite{Parisi:1983ae},
the noise-to-signal ratio $R_{NS}(t)$ of pion energy $E_\pi(\mathbf{p}=\tfrac{2\pi}{L}\mathbf{n})$
grows exponentially as
$
R_{NS}(t) \propto \frac{1}{\sqrt{N_{c}}} \exp {(\sqrt{m_{\pi}^2 + \tfrac{4\pi^2}{L^2}\mathbf{n}^2}-m_\pi)t},
$
where $N_{c}$ is the number of gauge configurations.
Consequently, for a given momentum $\mathbf{p}$,
one of the most efficient ways to improve the statistics
is to choose lattice ensembles with higher lattice spatial dimension $L$ (see more discussion in the Appendix),
which are explored by RQCD~\cite{Bali:2015gji}.

It is economical to perform lattice studies using staggered fermions
compared to using other discretizations;
this permits lattice examinations with the larger lattice spatial dimensions $L$
or smaller quark masses within limited computer resources.
For this reason, we first use staggered fermions to examine $\rho$ resonance parameters,
and then carry out lattice calculations on MILC lattice ensembles
with Asqtad-improved staggered sea quarks
(we use two ensembles with $L=40$ and one ensemble with $L=64$).
This not only allows us to measure the pion energy
for higher momenta with high quality,
but also enables us to study $\pi\pi$ scattering for the moving frame
with total momentum ${\mathbf P}=(2\pi/L)({\mathbf e}_1+{\mathbf e}_2+{\mathbf e}_3)$
and ${\mathbf P}=(2\pi/L)2{\mathbf e}_3$,
which are explored by HSC~\cite{Dudek:2012xn}.

To map out the resonance region efficiently, for each lattice ensemble
we study the $I=1$ $\pi\pi$ system with five Lorentz frames, one CMF and four MFs.
The first moving frame is implemented with total momentum ${\mathbf P}=(2\pi/L){\mathbf e}_3$ (MF1),
the second moving frame with ${\mathbf P}=(2\pi/L)({\mathbf e}_1+{\mathbf e}_2)$ (MF2),
and the third moving frame with ${\mathbf P}=(2\pi/L)({\mathbf e}_1+{\mathbf e}_2+{\mathbf e}_3)$ (MF3),
where the ${\mathbf e}_i$ is  a unit vector in the spatial direction $i$.
For the large lattice space (i.e., $L\ge 32$), we also use a fourth moving frame
with ${\mathbf P}=(2\pi/L)2{\mathbf e}_3$ (MF4).
For a CMF, we extract the $p$-wave scattering phase only from the energy levels of the ground state;
for each of the MFs, we extract them from the energy levels of the ground state and the first excited state:
consequently, we can obtain the scattering phases at seven or nine energies for six MILC lattice ensembles.
We will find that usually at least four energies are calculated
for the  $p$-wave $I=1$ $\pi\pi$ scattering phases,
which either lie in or are in the vicinity of the resonance range $[m_\rho-\Gamma_\rho, m_\rho+\Gamma_\rho]$.

The lattice ensemble parameters of the MILC gauge configurations have been reliably determined by
the MILC Collaboration~\cite{Bernard:2010fr,Bazavov:2009bb}.
Our lattice simulation used six pion masses ranging from $346$ to $176$~MeV,
ensuring that the physical kinematics
for the $\rho$-meson decay, $m_\pi/m_\rho<0.5$, is satisfied.
Moreover, the computation of $\rho$ resonance parameters at six lattice ensembles
allows us, following the ETMC Collaboration~\cite{Feng:2010es},
to investigate the pion mass dependence of
the resonance mass and decay width and, hence, to
reliably perform a chiral extrapolation to the physical point.
Additionally, our numerical calculations of the $\pi\pi$ correlators
are for the first time calculated with the moving-wall source,
which allows us to obtain results with high statistics.
Moreover, according to the discussion in the Appendix,
the usage of lattice ensembles with relatively large $L$
and the summations of the $\rho$ correlator over all the even time slices
and the $\pi\pi$ correlator over all the time slices
also significantly improved the signals of the corresponding correlators.

This article is organized as follows.
In Sec.~\ref{Sec:Methods}, we elaborate on our calculation method.
Our concrete lattice calculations are provided in Sec.~\ref{sec:latticeCal}.
We deliver our lattice results in Sec.~\ref{Sec:Results},
provide analysis in Sec.~\ref{Sec:analysis},
and reach our conclusions and outlooks in Sec.~\ref{Sec:Conclusions}.
Discussions of the  noise-to-signal ratio of correlator
are left to the Appendix.

\newpage
\section{Finite-volume methods}
\label{Sec:Methods}
In the present study, we will examine the neutral $\rho$-meson decay
into a pair of pions in the $p$-wave state,
and concentrate on $\pi\pi$ system with
the isospin representation of $(I,I_z)=(1,0)$.
We restrict ourselves to the overall
momenta $\mathbf{P}=[0,0,0]$, $[0,0,1]$, $[1,1,0]$, $[1,1,1]$ and $[0,0,2]$.\footnote{
The momentum is written in units of $\tfrac{2\pi}{L}$.
For easy notation, in some places of this paper, the square braces
are adopted to suggest a suppression of the dimensional factor,
to be specific, $\mathbf{P}=[0,0,0]$ denotes a momentum of $(0,0,0)\tfrac{2\pi}{L}$.
}

\subsection{Center-of-mass frame}
In the center-of-mass frame, the energy levels of two free pions are provided by
$$
E = 2\sqrt{m_\pi^2+ |{\mathbf p}|^2} ,
$$
where ${\mathbf p}=\tfrac{2\pi}{L}{\mathbf n}$, and
${\mathbf n}\in \mathbb{Z}^3$.
For the lattice ensembles with enough large $L$ and small pion masses,
the lowest energy $E$ for ${\mathbf n} \ne 0$ [e.g., $\mathbf{n}=(1,0,0)$]
is usually in the vicinity of  the lattice-measured $\rho$ mass $m_{\rho}$.
Therefore, we will pay special attention to the ${\mathbf n} = (1,0,0)$ case.
In fact, we indeed calculate the energy levels for the ${\mathbf n} = (1,1,0)$ and ${\mathbf n} = (1,1,1)$ cases.
Unfortunately, almost all of these energy levels either turned out to
be beyond $4\pi$ threshold or the relevant signals were not good enough.
We should remark at this point that the finite-volume methods
are only valid for the elastic scattering;
consequently, we are only interested in the energy levels
of the $\pi\pi$ system in the elastic region $2m_\pi < E < 4m_\pi$.

In the presence of the interaction between two pions,
the energy levels of the $\pi\pi$ system are displaced
by the hadronic interaction from $E$ to $\overline{E}$,
$$
\overline{E} = 2\sqrt{m_\pi^2 + k^2} ,
\quad k=\frac{2\pi}{L}q ,
$$
where the dimensionless scattering momentum $q \in \mathbb{R}$.
These energy levels transform
as the irreducible representation $T_1^-$ under the cubic group $O_h$.
The L\"uscher formula links the energy $\overline{E}$
to the $p$-wave $\pi\pi$ scattering phase
$\delta_1$~\cite{Luscher:1990ux,Luscher:1990ck},
\begin{equation}
\label{eq:CMF}
\tan\delta_1(k)=\frac{\pi^{3/2}q}{\mathcal{Z}_{00}(1;q^2)} ,
\end{equation}
where the zeta function is formally defined by
\begin{equation}
\label{eq:Zeta00_CM}
\mathcal{Z}_{00}(s;q^2)=\frac{1}{\sqrt{4\pi}}
\sum_{{\mathbf n}\in\mathbb{Z}^3} \frac{1}{\left(|{\mathbf n}|^2-q^2\right)^s}  .
\end{equation}
The zeta function $\mathcal{Z}_{00}(s;q^2)$
can be efficiently evaluated by the method described in Ref.~\cite{Yamazaki:2004qb}.
We notice an equivalent L\"uscher formula has been recently developed in Ref.~\cite{Doring:2011vk}.

\subsection{Moving frame}
Using a moving frame with nonzero total momentum
${\mathbf P}=(2\pi/L){\mathbf d}$, ${\mathbf d}\in\mathbb{Z}^3$,
the energy levels of two free pions are expressed by
$$
E_{MF} = \sqrt{m_\pi^2 + |{\mathbf p}_1|^2} +
         \sqrt{m_\pi^2 + |{\mathbf p}_2|^2} ,
$$
where ${\mathbf p}_1$, ${\mathbf p}_2$ denote the three-momenta of the pions,
which obey the periodic boundary condition,
${\mathbf p}_1=\frac{2\pi}{L}{\mathbf n}_1$,
${\mathbf p}_2=\frac{2\pi}{L}{\mathbf n}_2$,
${\mathbf n}_1,{\mathbf n}_2\in \mathbb{Z}^3$,
and total momentum ${\mathbf P}$ is
$
{\mathbf P} = {\mathbf p}_1 + {\mathbf p}_2$~\cite{Rummukainen:1995vs}.

In the presence of an interaction between two pions,
the energy $E_{CM}$  is
\begin{equation}
\label{eq:ECM_q}
E_{CM} =
2\sqrt{m_\pi^2 + p^{*2}} ,
\quad p^{*} = \frac{2\pi}{L} q ,
\end{equation}
where the dimensionless momentum $q \in \mathbb{R}$, $p^*=| {\mathbf p}^*|$,
and ${\mathbf p}^*$ are quantized to the values
$
{\mathbf p}^* =\frac{2\pi}{L}{\mathbf r},
{\mathbf r} \in P_{\mathbf d} ,
$
and the set $P_{\mathbf d}$  is
\begin{equation}
\label{eq:set_Pd_MF}
P_{\mathbf d} = \left\{ {\mathbf r} \left|  {\mathbf r} = \vec{\gamma}^{-1}
\left[{\mathbf n}+ \frac{\mathbf d}{2}
\right], \right.  {\mathbf n}\in\mathbb{Z}^3 \right\} ,
\end{equation}
where $\vec{\gamma}^{-1}$ is the inverse Lorentz transformation
operating in the direction of the center-of-mass velocity ${\mathbf v}$,
$
\vec{\gamma}^{-1}{\mathbf p} =
\gamma^{-1}{\mathbf p}_{\parallel}+{\mathbf p}_{\perp} ,
$
where ${\mathbf p}_{\parallel}$ and ${\mathbf p}_{\perp}$ are
the ingredients of ${\mathbf p}$ parallel
and perpendicular to ${\mathbf v}$, respectively.
Using the Lorentz transformation,
the energy $E_{CM}$ is connected to the $E_{MF}$  through
$E_{CM}^2 = E_{MF}^2-{\mathbf P}^2.$

The scattering phase shifts are expressed in terms of the
generalized zeta function
\begin{equation}
\mathcal{Z}^{ \mathbf d }_{\ell m } ( s ; q^2 ) =
\sum_{ {\mathbf r} \in P_{\mathbf d} } \frac{ r^\ell Y_{\ell m}(\Omega_r)  }{( r^2 - q^2 )^s }  ,
\label{zetafunction_MF}
\end{equation}
where the set $P_{\mathbf d}$  is defined in Eq.~(\ref{eq:set_Pd_MF}),
the $Y_{\ell m}$ are the spherical harmonic functions,
and $\Omega_r$ stands for the solid angle parameters $(\theta, \phi)$ of $\mathbf{r}$
in spherical coordinates.

The first moving frame (MF1) is taken with ${\mathbf d}={\mathbf e}_3$,
and the energy levels of the $\pi\pi$ system transform under the tetragonal group $D_{4h}$.
The irreducible representations $A_2^-$ and $E^-$ are associated with
the $p$-wave $\pi\pi$ scattering states in a torus.
In this work, we are only interested in the $A_2^-$ sector due to limited computer resources;
the energy levels $\overline{E}$ are linked to
the $p$-wave $\pi\pi$ scattering phase shift $\delta_1$ with the Rummukainen-Gottlieb formula
for the $A_2^-$  representation~\cite{Rummukainen:1995vs},
\begin{eqnarray}
\label{eq:Luscher_MF1A2}
\tan\delta_1(q)  &=&
\frac{\gamma\pi^{3/2}q}{\mathcal{Z}_{00}^{\mathbf d}(1;q^2)
+\frac{2}{\sqrt{5}}q^{-2} \mathcal{Z}_{20}^{\mathbf d}(1;q^2)} ,
\end{eqnarray}
where the higher scattering phase shifts
$\delta_\ell (\ell\ge3)$ are ignored,
and the dimensionless center-of-mass scattering momentum $q$ is calculated
from the lattice-measured energies of the $\pi\pi$ system through Eq.~(\ref{eq:ECM_q}).
The boost factor $\gamma$ is calculated by $\gamma=E_{MF}/E_{CM}.$

We implemented the second moving frame (MF2) with ${\mathbf d}={\mathbf e}_1+{\mathbf e}_2$,
and the corresponding energy levels of the $\pi\pi$ system  transform
under the orthorhombic group $D_{2h}$.
The irreducible representations
$A_1^-$, $B_1^-$, and $B_2^-$ occur for $p$-wave $\pi\pi$ scattering states in a torus.
In this work, we concentrate on the $B_1^-$ sector;
the corresponding finite-size formula for $B_1^-$  representation is given by~\cite{Rummukainen:1995vs,Feng:2010es,Lang:2011mn,Gockeler:2012yj}
\begin{eqnarray}
\label{eq:MF2}
\hspace{-0.8cm}\tan\delta_1(k)&=& \cr
&&\hspace{-1.8cm} \frac{\gamma\pi^{3/2}q}{\mathcal{Z}_{00}^{\mathbf d}(1;q^2)
-\frac{1}{\sqrt{5}}q^{-2}\mathcal{Z}_{20}^{\mathbf d}(1;q^2)
-i\sqrt{\frac{6}{5}}q^{-2}\mathcal{Z}_{22}^{\mathbf d}(1;q^2)} ,
\end{eqnarray}
where the higher scattering phase shifts $\delta_\ell (\ell\ge3)$ are ignored.

In order to acquire more eigenenergies in the resonance region,
we considered the third moving frame (MF3) with
${\mathbf d}={\mathbf e}_1+{\mathbf e}_2 + {\mathbf e}_3$.
The corresponding energy eigenstates transform under the orthorhombic group $D_{3d}$.
The irreducible representations
$A_2^-$, and $E^-$ occur for the $p$-wave $\pi\pi$ scattering states in a torus.
Here we are only interested in the $A_2^-$ sector;
the corresponding finite-size formula for MF3
with $A_2^-$  representation is provided by~\cite{Dudek:2012xn,Gockeler:2012yj}
\begin{eqnarray}
\label{eq:MF3}
\hspace{-0.5cm}\cot\delta_1(k)&=&
\frac{1}{\gamma\pi^{3/2}q} \left\{ \mathcal{Z}_{00}^{\mathbf d}(1;q^2)-
i\sqrt{\frac{8}{15}}\frac{1}{q^{2}}\mathcal{Z}_{22}^{\mathbf d}(1;q^2)\right. \cr
&-& \left.
\sqrt{\frac{8}{15}}\frac{1}{q^{2}}\left[{\rm Re}\mathcal{Z}_{21}^{\mathbf d}(1;q^2)+
{\rm Im}\mathcal{Z}_{21}^{\mathbf d}(1;q^2)\right] \right\},
\end{eqnarray}
where we overlook the higher scattering phase shifts
$\delta_\ell(\ell\ge3)$ as well.

\begin{table*}[t]
\caption{
Summary of the irreducible representations for the ground and first excited states with the isospin $(I,I_z)=(1,0)$, where ${\mathbf P}$ denotes total momentum,
$g$ gives the rotational group in each frame and $\Gamma$ shows the relevant irreducible representation.
The two-pion operators ${\cal O}_{\pi\pi}$ and rho operators ${\cal O}_{\rho}$  are listed in Column $5$ and $6$, respectively.
The vectors in parentheses behind $\pi$ and $\rho$ represent
the momenta of the two-pion state and the $\rho$ meson in units of $2\pi/L$, respectively.
}
\label{table:calc_rep}
\begin{ruledtabular}
\begin{tabular}{lcll cc}
Frame & ${\mathbf P}$ & $g$  & $\Gamma$ & ${\cal O}_{\pi\pi}$ & ${\cal O}_{\rho}$ \\
\hline
CMF  & $[0,0,0]$  & $O_h$ & ${T}_{1}^{-}$ & $\pi(0,0,1)\pi(0,0,-1)$
& $(\rho_{1}+\rho_{2}+\rho_{3})(0,0,0)$ \\
\hline
MF1   & $[0,0,1]$  & $D_{4h}$ & ${A}_{2}^{-}$ & $\pi(0,0,1)\pi(0,0,0)$   & $\rho_{3}(0,0,1)$ \\
\hline
MF2  & $[1,1,0]$  & $D_{2h}$    & ${B}_{1}^{-}$ & $\pi(1,1,0)\pi(0,0,0)$ & $(\rho_{1}+\rho_{2})(1,1,0)$ \\
\hline
MF3  & $[1,1,1]$  & $D_{3h}$ & ${A}_{2}^{-}$ & $\pi(1,1,1)\pi(0,0,0)$
     & $(\rho_{1}+\rho_{2}+\rho_{3})(1,1,1)$ \\
\hline
MF4  & $[0,0,2]$  & $D_{4h}$ & ${A}_{2}^{-}$ & $\pi(0,0,2)\pi(0,0,0)$ & $\rho_{3}(0,0,2)$
\end{tabular}
\end{ruledtabular}
\end{table*}

For large lattice ensembles ( i.e., $L\ge32$), the fourth moving frame (MF4)
with ${\mathbf d}={2\mathbf e}_3$ is also considered,
and the energy levels of $\pi\pi$ system transform under the tetragonal group $D_{4h}$.
The irreducible representations $A_2^-$ and $E^-$ are associated
with the $p$-wave $\pi\pi$ scattering states in a torus.
We here concentrate on the $A_2^-$ sector.
The energy levels $\overline{E}$ linked to
the $p$-wave $\pi\pi$ scattering phase $\delta_1$ with the Rummukainen-Gottlieb formula
for the $A_2^-$  representation can be calculated by Eq.~(\ref{eq:Luscher_MF1A2}).

The calculation method of zeta functions $\mathcal{Z}_{00}^{{\mathbf d}} (1; q^2)$,
$\mathcal{Z}_{21}^{{\mathbf d}} (1; q^2)$, and $\mathcal{Z}_{22}^{{\mathbf d}} (1; q^2)$
is elaborated in Appendix A of Ref.~\cite{Yamazaki:2004qb},
where we also gave its extensions in the two-particle system with arbitrary masses~\cite{Fu:2012gf}.
In this work, we are particularly interested in a MF3,
with one pion at rest and one pion with momentum
${\mathbf p} = (2\pi / L) ({\mathbf e}_1+{\mathbf e}_2+{\mathbf e}_3)$,
and a MF4, with one pion at rest and one pion with momentum
${\mathbf p} = (4\pi / L) {\mathbf e}_3$.
For our concrete calculations, we found that the relevant
scattering phases are usually calculated at energies which are more efficiently used to
directly mark out the resonance region.

In this work, we only calculate the scattering phase
of the ground state for the ${\bf T}^{-}_1$ representation,
since the relevant eigenenergies are expected to be much smaller
than those of the excited states~\cite{Aoki:2011yj}.
For the ${\bf A}_{2}^{-}$ and the ${\bf B}_{1}^{-}$ representations,
we will also calculate the scattering phase shift
for the first excited state.
The relevant representations for the ground and the first excited states
with the isospin $(I,I_z)=(1,0)$
are summarized in Table~\ref{table:calc_rep}.

\subsection{Variational analysis}
\label{SubSec:Correlation_matrix }

In order to extract the energy eigenvalues of the lower two states for the ${\bf A}_{2}^{-}$
and the ${\bf B}_{1}^{-}$ representations discussed in Sec.~\ref{Sec:Methods}$-$i.e., $\overline{E}_n$ ($n=1,2$)$-$ the state-of-the-art variational method~\cite{Luscher:1990ck} is exploited for Wilson fermions.
Moreover, corrections to the true energy levels are discussed in detail when the energies
are extracted from the generalized eigenvalues~\cite{Blossier:2009kd}.
These methods can readily be applied to staggered fermions with a small alteration~\cite{DeTar:2014gla}.
In practice, we employ a two-dimensional variational basis
and build the correlation function matrix,
\begin{eqnarray}
\hspace{-0.3cm}C(t) &= & \cr
&&\hspace{-1.2cm}\left(
\begin{array}{ll}
\langle 0 | {\cal O}_{\pi\pi}^\dag(\mathbf{p}, t)  {\cal O}_{\pi\pi}(\mathbf{p}, 0)  | 0 \rangle &
\langle 0 | {\cal O}_{\pi\pi}^\dag(\mathbf{p}, t)  {\cal O}_{\rho}(\mathbf{p}, 0) | 0 \rangle
\vspace{0.3cm} \\
\langle 0 | {\cal O}_{\rho}^\dag(\mathbf{p}, t) {\cal O}_{\pi\pi}(\mathbf{p}, 0) | 0 \rangle &
\langle 0 | {\cal O}_{\rho}^\dag(\mathbf{p}, t) {\cal O}_{\rho}(\mathbf{p}, 0)| 0 \rangle
\end{array} \right), \cr
&&
\label{eq:CorrMat}
\end{eqnarray}
where ${\cal O}_{\rho}$ is an interpolator
for the neutral $\rho$ meson with the specified momentum ${\mathbf p} $
and the polarization vector parallel to the $\rho$ momentum ${\mathbf p}$,
and ${\cal O}_{\pi\pi}$ is an interpolator for the $\pi\pi$ system
with the given total momentum  ${\mathbf P} = {\mathbf p}$.

\begin{figure*}[htp!]
\includegraphics[width=16cm,clip]{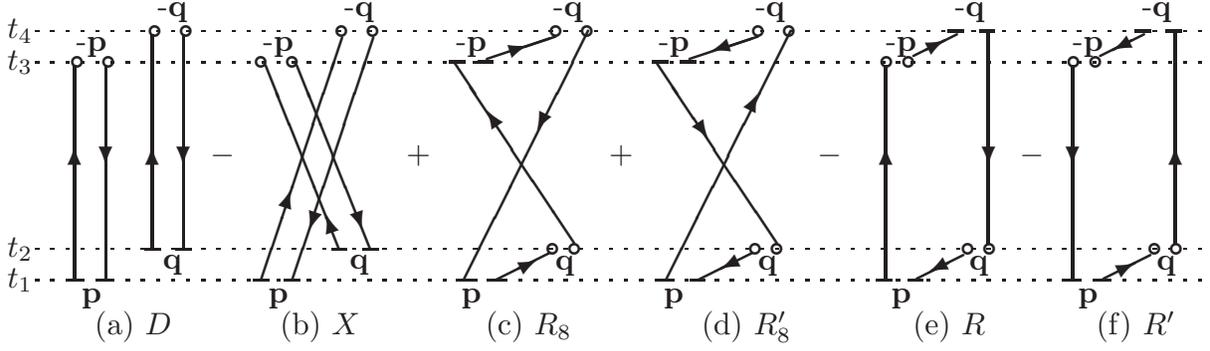}
\caption{ \label{fig:diagram}
(color online).
Quark-link diagrams contributing to the  $I=1$ $\pi\pi$ four-point functions.
Short black bars stand for the wall sources.
Open circles are sinks for local pion operators.
The time flows upward in the diagrams.
The pion operators are given with a momentum specified in the diagram.
}
\end{figure*}

\subsubsection{$\pi\pi$ sector}
In this section, the original definitions and notations
are employed to review the basic formula for the lattice calculation
of the $p$-wave scattering phase of the $I=1$ $\pi\pi$ system
enclosed in a cubic torus~\cite{Kuramashi:1993ka,Fukugita:1994ve}.
Let us concentrate on the scattering of two Nambu-Goldstone pions
in the Kogut-Susskind staggered fermion formalism.

We build the $\pi\pi$ interpolator with the isospin representation
$(I,I_z)=(1, 0)$ as~\cite{Aoki:2007rd,Frison:2010ws}
\begin{eqnarray}
\label{EQ:op_pipi}
\hspace{-1.5cm} {\cal O}_{\pi\pi}^{I=1}({\mathbf{p,q}}; t) &=& \frac{1}{\sqrt{2}}
\big( \pi^+({\mathbf{q}},t) \pi^-({\mathbf{p}}, t+1)  \cr
&&    -\pi^+({\mathbf{p}},t) \pi^-({\mathbf{q}}, t+1)  \big)  ,
\end{eqnarray}
where the pion momenta ${\mathbf{p}} \ne {\mathbf{q}}$
and the total momentum of the $\pi\pi$ system  ${\mathbf{P}} = {\mathbf{p}} + {\mathbf{q}}$.
In order to avoid the complicated Fierz rearrangement of quark lines~\cite{Fukugita:1994ve},
we choose creation operators at time slices that are different by one lattice time spacing.

The operator that creates a single pion with nonzero momentum $\mathbf{p}$
from the vacuum is obtained by the Fourier transform
$
{\cal O}_{\pi}({\mathbf{p}},t) = \sum_{\mathbf{x}}
e^{ i{\mathbf p} \cdot {\mathbf{x}} }  {\cal O}_{\pi}({\mathbf{x}},t),
$
where the pion interpolators are denoted by
${\cal O}_{\pi^+}({\mathbf{x}},t) =
- \overline{d}({\mathbf{x}},t)\gamma_5 u({\mathbf{x}},t)$,
${\cal O}_{\pi^-}({\mathbf{x}},t) =
  \overline{u}({\mathbf{x}},t)\gamma_5 d({\mathbf{x}},t).
$

In the present work, we will concentrate on the five irreducible
representations, $T_1^-$, $A_2^-$, $B_1^{-}$, $A_2^-$, and $A_2^-$,  for the
CMF, MF1, MF2, MF3, and MF4, respectively.
In practice,  for  CMF,
the $\pi\pi$ interpolator is implemented
with ${\mathbf{q}} = -{\mathbf{p}}$ and $\mathbf{p}$= $[0,0,1]$.
For the $A_2^-$ and $B_1^{-}$ irreducible representations of four MFs, the $\pi\pi$ interpolators are
all taken with ${\mathbf{q}} = {\mathbf{0}}$,
and we calculate at four momenta, $\mathbf{p}$= $[0,0,1]$, $[1,1,0]$, $[1,1,1]$  and $[0,0,2]$
for each of the four moving frames, respectively.

In the isospin limit, topologically only six quark-line diagrams
contribute to $I=1$ $\pi\pi $ scattering amplitudes,
which are schematically illustrated in Fig.~1 of Ref.~\cite{Aoki:2007rd}.
These diagrams are also elucidated in Fig.~\ref{fig:diagram},
where four pions are placed at lattice points $x_1, x_2, x_3$, and $x_4$,
respectively [$x_1 \equiv ({\mathbf{x}}_1,t_1)$, etc.].
We usually label these diagrams  as $D$, $X$, $R_8$, $R_8^\prime$, $R$,
and $R^\prime$ diagrams, respectively.
The second diagram is also a kind of direct diagram;
we call it $X$ since its shape looks like an ``$X$''.
The shape of the third diagram looks like the number $8$; thus, we use ``$R_8$'' to identify it,
which is also a kind of the rectangular diagram. 
The superscript prime in $R$ and $R_8$ indicates the corresponding counterclockwise partners.

The moving-wall source technique was initially introduced by Kuramashi {\it et al.}~\cite{Kuramashi:1993ka,Fukugita:1994ve}
to study the $I=0, 2$ $\pi\pi$ scattering in the center-of-mass frame.
Recently, we further extended this technique to the two-particle system
with nonzero momenta to tentatively investigate
the scalar $\kappa$, $\sigma$,  and vector $K^\star(892)$ meson decays~\cite{Fu:2012gf}.
In the present study, we use this technique to calculate
the $I=1$ $\pi\pi $ scattering amplitudes by computing each $T$ quark propagator
corresponding to the moving-wall source at all the time slices~\cite{Kuramashi:1993ka,Fukugita:1994ve,Fu:2012gf},
$$
\sum_{n'}D_{n,n'}G_t^{\mathbf p}(n') = \sum_{\mathbf{x}}
\delta_{n,({\mathbf{x}},t)}, \quad 0 \leq t \leq T-1 ,
$$
where $D$ is the Dirac quark matrix,
the subscript $t$ in the quark propagator $G$ indicates
the temporal position of the wall source~\cite{Kuramashi:1993ka,Fukugita:1994ve,Fu:2012gf},
and the superscript ${\mathbf p}$ in $G$
suggests that for the specified momentum $\mathbf{p}$,
we select an up-quark source or sink with $e^{i{\mathbf p} \cdot {\mathbf{x}}}$
(while an up-antiquark source or sink with $1$)
on each lattice site ${\mathbf{x}}$ for the pion creation operator~\cite{Bernard:2001av,Aubin:2004wf,Fu:2012gf}.
The associations of the quark propagators $G_t^{\mathbf p}(n)$
exploiting the $I=1$ $\pi\pi$ four-point correlation functions are schematically illustrated in Fig.~\ref{fig:diagram}~\cite{Kuramashi:1993ka,Fukugita:1994ve,Fu:2012gf}.
In terms of the quark propagators $G_t^{\mathbf p}(n)$, the $D$, $X$, $R_8$, $R_8^\prime$, $R$
and $R^\prime$ quark-line diagrams can be represented as
\begin{widetext}
\begin{eqnarray}
\label{eq:dcr}
C^D_{\pi\pi}({\mathbf{p,q}}; t_4,t_3,t_2,t_1)  &=&
\sum_{\mathbf{x}_3}\sum_{\mathbf{x}_4}
e^{ -i({\mathbf p} \cdot {\mathbf{x}}_3 + {\mathbf q} \cdot {\mathbf{x}}_4 )}
\langle \mbox{Tr}
[G_{t_1}^{\dag}({\mathbf{x}}_3,t_3) G_{t_1}^{\mathbf p}({\mathbf{x}}_3,t_3)] \,
\mbox{Tr}
[G_{t_2}^{\dag}({\mathbf{x}}_4,t_4) G_{t_2}^{\mathbf q}({\mathbf{x}}_4,t_4)] \rangle,\cr
C^{X}_{\pi\pi}({\mathbf{p,q}}; t_4,t_3,t_2,t_1) &=&
\sum_{\mathbf{x}_3}\sum_{\mathbf{x}_4}
e^{ -i({\mathbf p} \cdot {\mathbf{x}}_3 + {\mathbf q} \cdot {\mathbf{x}}_4 )}
\langle \mbox{Tr}
[G_{t_1}^{\dag}({\mathbf{x}}_4,t_4) G_{t_1}^{\mathbf p}({\mathbf{x}}_4,t_4)] \,
\mbox{Tr}
[G_{t_2}^{\dag}({\mathbf{x}}_3,t_3) G_{t_2}^{\mathbf q}({\mathbf{x}}_3,t_3)] \rangle,\cr
C^{R_8}_{\pi\pi}({\mathbf{p,q}}; t_4,t_3,t_2,t_1) &=&
\sum_{\mathbf{x}_2}\sum_{\mathbf{x}_4}
e^{ i{\mathbf q} \cdot ( {\mathbf{x}}_2 - {\mathbf{x}}_4) }
\langle \mbox{Tr}
[ G_{t_1}^{ \mathbf p}({\mathbf{x}}_2, t_2) G_{t_3}^\dag({\mathbf{x}}_2, t_2)
  G_{t_3}^{-\mathbf p}({\mathbf{x}}_4, t_4) G_{t_1}^\dag({\mathbf{x}}_4, t_4)
] \rangle,\cr
C^{R_8^\prime}_{\pi\pi}({\mathbf{p,q}}; t_4,t_3,t_2,t_1) &=&
\sum_{\mathbf{x}_2}\sum_{\mathbf{x}_4}
e^{ i{\mathbf q} \cdot ( {\mathbf{x}}_2 - {\mathbf{x}}_4) }
\langle \mbox{Tr}
[ G_{t_3}^{-\mathbf p}({\mathbf{x}}_2, t_2) G_{t_1}^\dag({\mathbf{x}}_2, t_2)
  G_{t_1}^{ \mathbf p}({\mathbf{x}}_4, t_4) G_{t_3}^\dag({\mathbf{x}}_4, t_4)
] \rangle,\cr
C^{R}_{\pi\pi}({\mathbf{p,q}}; t_4,t_3,t_2,t_1) &=&
\sum_{\mathbf{x}_2}\sum_{\mathbf{x}_3}
e^{ i({\mathbf q} \cdot {\mathbf{x}}_2 - {\mathbf p} \cdot {\mathbf{x}}_3)}
\langle \mbox{Tr}
[G_{t_1}^{\mathbf p}({\mathbf{x}}_3, t_3) G_{t_4}^{\dag}({\mathbf{x}}_3, t_3)
 G_{t_4}^{-\mathbf q}({\mathbf{x}}_2, t_2) G_{t_1}^{\dag}({\mathbf{x}}_2, t_2)
] \rangle,\cr
C^{R^\prime}_{\pi\pi}({\mathbf{p,q}}; t_4,t_3,t_2,t_1) &=&
\sum_{\mathbf{x}_2}\sum_{\mathbf{x}_3}
e^{ i({\mathbf q} \cdot {\mathbf{x}}_2 - {\mathbf p} \cdot {\mathbf{x}}_3)}
\langle \mbox{Tr}
[G_{t_1}^{\mathbf p}({\mathbf{x}}_2, t_2)  G_{t_4}^{\dag}({\mathbf{x}}_2, t_2)
 G_{t_4}^{-\mathbf q}({\mathbf{x}}_3, t_3) G_{t_1}^{\dag}({\mathbf{x}}_3, t_3)
] \rangle,
\end{eqnarray}
\end{widetext}
where the traces are carried out over the color index,
and the $\gamma_5$-Hermiticity nature of the light quark propagator $G$, {\it i.e.},
$G(t,t')^\dagger=\gamma_5 G(t',t)\gamma_5$, has been applied~\cite{Kuramashi:1993ka,Fukugita:1994ve}.

According to the discussions in Ref.~\cite{Pelissier:2012pi}, in the isospin limit,
the real parts of the third and fourth quark-line diagrams in Fig.~\ref{fig:diagram}
have the same values, while the corresponding imaginary parts have the same magnitudes as well,
just with the opposite sign
(likewise for the fifth and sixth quark-line diagrams).\footnote{
This is true for the average with respect to the gauge configurations~\cite{Pelissier:2012pi}.
Note that this is true even for calculations at a single gauge configuration
because the moving-wall source technique, a nonstochastic method, is used.
This can be readily verified from the analytical expression in Eq.~(\ref{eq:dcr}),
where the random numbers are not used in these expressions.
}
Therefore, the value of the $I=1$ $\pi\pi$ four-point correlation function is purely real.
Consequently, only four quark-line diagrams ($D$, $X$, $R_8$, and $R$)
are needed to calculate the $I=1$ $\pi\pi$ four-point correlation function, namely,
\begin{eqnarray}
\label{EQ:phy_I12_32}
\hspace{-2cm} C_{\pi\pi}({{\mathbf p},{\mathbf q}}; t)
&\equiv&
\left\langle {\cal O}_{\pi\pi}({{\mathbf p},{\mathbf q}}; t) |
             {\cal O}_{\pi\pi}({{\mathbf p},{\mathbf q}}; 0) \right\rangle \cr
\hspace{-2cm} &=&
D - X + 2 N_f R_8 - 2 N_f R ,
\end{eqnarray}
where the staggered-flavor factor $N_f$ should be inserted into
the rectangular diagrams ($R$ and $R_8$) to amend for
the additional factor $N_f$ in the valence fermion loops~\cite{Sharpe:1992pp}.
We should remark at this point that the fourth-root recipes are supposed to
correctly recover the right continuum limit of QCD~\cite{Durr:2004as}.

\subsubsection{$\rho$ sector}
In principle, we can measure the propagators for two local $\rho$ mesons,
$\gamma_i \otimes \gamma_i$ (VT)  and
$\gamma_0 \gamma_i \otimes \gamma_0\gamma_i$ (PV)~\cite{Bernard:2001av,Aubin:2004wf}.
Nonetheless, we merely present lattice results for the local
VT $\rho$  meson because it gives more stable signals~\cite{Bernard:2001av,Aubin:2004wf}.
Additionally, the numerical calculation of the
three-point correlation function $\rho \to \pi\pi$ is rather simple to compute
if the local VT $\rho$ interpolator is used.
Therefore, we only employ an interpolator
with isospin $I=1$ and $J^{P}=1^{-}$ at source and sink,
$$
{\cal O}(x)  \equiv \sum_{c}
\frac{1}{\sqrt{2}} \left\{
u_c(x) \gamma_i \otimes \gamma_i \bar{u}_c( x ) -
d_c(x) \gamma_i \otimes \gamma_i \bar{d}_c( x )
\right\},
$$
where $c$ is the color index,
and the subscript $i$ in $\gamma_i$ indicates the polarizations
of the $\rho$ vector current.

In the isospin limit, the disconnected quark-line diagrams
for the $\rho$ meson are nicely canceled out;
consequently, the correlator for the neutral $\rho$  meson
in the momentum ${\mathbf p}$ state is solely computed by the connected diagram
\begin{eqnarray}
C_{\rho}({\mathbf{p} }, t) &=& \sum_{\mathbf{x}} \sum_{ a, b}
e^{i{\mathbf p} \cdot {\mathbf x} }
\langle u_b( {\mathbf{x} }, t) \gamma_i \otimes \gamma_i \bar u_b ({\mathbf{x}}, t)  \cr
&&\times
u_{a}({\mathbf 0}, 0) \gamma_i \otimes \gamma_i \bar u^{a}_{g }({\mathbf 0}, 0)  \rangle ,
\nonumber
\end{eqnarray}
where ${\mathbf 0}, \mathbf{x}$ are lattice spatial points of
$\rho$ states at the source and sink, respectively.
In practice, we use the wall-source and point-sink interpolators
to efficiently reduce the overlap with the excited states~\cite{Aubin:2004fs}.

We fit the $\rho$ correlator with the physical model as
\begin{eqnarray}
\label{eq:kfit}
C_{\rho}(t)  &=&
A \cosh\left[m\left(t-\tfrac{T}{2}\right)\right] \cr
&+& A^{\prime}(-1)^{t+1} \cosh\left[m^\prime \left(t-\tfrac{T}{2}\right)\right] ,
\nonumber
\end{eqnarray}
where $A$ and $A^\prime$ are two overlap amplitudes,
where only one mass is taken with
each parity~\cite{Golterman:1985dz,Bernard:2001av,Aubin:2004wf},
and the oscillating parity partner is the $p$-wave meson with $J^P=1^+$.

\subsubsection{Off-diagonal sector}
\begin{figure}[htp!]
\includegraphics[width=8.5cm,clip]{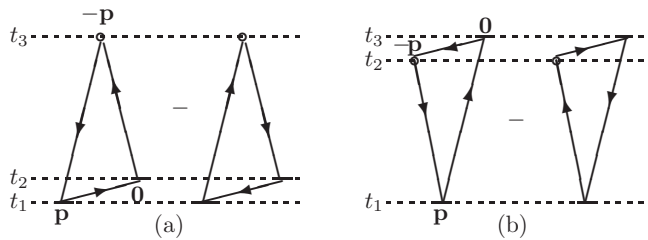}
\caption{ \label{fig:3diagram}
(color online). Quark-link diagrams contributing to $\pi\pi \to \rho$ and
$\rho \to \pi\pi$ three-point correlation functions.
Short black bars indicate the wall sources.
(a) Quark contractions of $\pi\pi \to \rho$,
where the open circle is the sink for $\rho$ operator.
(b) Quark contractions of $\rho \to \pi\pi$,
where the open circle is the sink for pion operator.
}
\end{figure}

When studying the resonance parameters of the $\rho$ meson~\cite{Loft:1988sy,Altmeyer:1995qx,McNeile:2002fh,Aoki:2007rd,Aoki:2011yj,Gockeler:2008kc,Feng:2010es,Frison:2010ws,Pelissier:2012pi,Dudek:2012xn,Lang:2011mn,Bulava:2015qjz,Guo:2016zos,Wilson:2015dqa,Bali:2015gji},
one chiefly employs the stochastic method
or its variants to measure the three-point function~\cite{Drummond:1982sk},
which are successfully recently measured with the moving-wall source technique~\cite{Fu:2012gf}.
To hinder the twisted color Fierz transformation of
the quark lines~\cite{Fukugita:1994ve},
we commonly choose $t_1 \ne t_2$.
In practice, we pick $t_1 =0, t_2=1$, and $t_3=t$
for the $\pi\pi \to \rho$ three-point correlation function,
and select $t_1 =0, t_2=t$, and $t_3=t+1$
for the $\rho \to \pi\pi$ three-point correlation function.
The quark-line diagrams corresponding to the $\rho \to \pi\pi$ and
$\pi \pi \to \rho$  are schematically illustrated
in Figs.~\ref{fig:3diagram}(a) and ~\ref{fig:3diagram}(b), respectively.

In practice, we employ an up-antiquark source with $1$ on each lattice site ${\mathbf{x}}$
for pion creation operator,
and an up-quark source with $e^{i{\mathbf{p}}\cdot{\mathbf{x}}}$
on each lattice site ${\mathbf{x}}$ for pion creation operator~\cite{Fu:2012gf}.
It should be worthwhile to stress that the imaginary part of the second diagram
for $\pi\pi \to \rho$ should have same magnitude but with the minus sign,
as that of the first diagram~\cite{Pelissier:2012pi}
(likewise for $\rho \to \pi\pi$).
As a consequence, the three-point diagrams are purely imaginary,
and only one quark-line diagram is required to calculate
each of the three-point correlation functions.
We then write each of the first diagrams for the $\rho \to \pi\pi$ and
$\pi\pi \to \rho$ quark-line diagrams in Fig.~\ref{fig:3diagram}
in terms of the light quark propagators $G$,

\begin{eqnarray}
\label{eq:dcr3}
C_{\pi\pi \to \rho} ({\mathbf p};t_3,t_2,t_1)
&&=\cr
&&\hspace{-3.0cm} \sum_{ {\mathbf{x}}_3, {\mathbf{x}}_2}
e^{ i {\mathbf p} \cdot {\mathbf{x}}_3 }
\langle  \mbox{Tr}
[G_{t_2}({\mathbf{x}}_3, t_3) \gamma_5
\, G_{t_1}^{\dag}({\mathbf{x}}_3, t_3) \gamma_i
          G_{t_1} ({\mathbf{x}}_2, t_2) ] \gamma_5  \rangle , \cr
C_{\rho \to \pi\pi}({\mathbf p};t_3,t_2,t_1)
&&= \cr
&&\hspace{-3.0cm}  \sum_{ {\mathbf{x}}_2, {\mathbf{x}}_3}
e^{ i {\mathbf p} \cdot {\mathbf{x}}_2 }
\langle  \mbox{Tr}
[G_{t_3}({\mathbf{x}}_2, t_2) \gamma_i \,
G_{t_1}^{\dag}({\mathbf{x}}_2, t_2) \gamma_5
 G_{t_1} ({\mathbf{x}}_3, t_3) ] \gamma_5 \rangle , \nonumber
\end{eqnarray}
where the trace is taken over the color index and the Dirac matrix
is used as an interpolator for the $i$th meson:
the $\gamma_5$ for the pseudoscalar meson and $\gamma_i$ for the vector $\rho$ meson,
where the subscript $i$ in the $\gamma_i$ indicates the polarization of
the $\rho$ vector current.
%

\subsubsection{ Extraction of energies }
\label{SubSec:Extraction of energies}
To map out the avoided level crossings between the $\rho$  resonance
and its decay products,
the variational method~\cite{Luscher:1990ck}
is applied to separate the ground state from the first excited state.
In practice, we calculate $2 \times 2 $ correlation function matrix
$C(t)$ denoted in~(\ref{eq:CorrMat}),
and construct a ratio of the correlation function matrices as
\begin{equation}
M(t,t_R) = C(t)  \, C^{-1}(t_R)  ,
\label{eq:M_def}
\end{equation}
with some reference time  $t_R$~\cite{Luscher:1990ck}
to extract two energy eigenvalues $\overline{E}_n$ ($n=1,2$),
which can be obtained by a cosh fit to two eigenvalues
$\lambda_n (t,t_R)$ ($n=1,2$) of the correlation matrix $M(t,t_R)$~\cite{Barkai:1985gy}
\begin{eqnarray}
\label{Eq:asy}
\lambda_n (t, t_R) &=&
       A_n \cosh\left[-E_n\left(t-\tfrac{T}{2}\right)\right] \cr
       &&+
(-1)^t B_n \cosh\left[-E_n^{\prime}\left(t-\tfrac{T}{2}\right)\right] .
\end{eqnarray}
Note that the relevant higher correction is discussed in Ref.~\cite{DeTar:2014gla}.
In practice, we will remove the ``wraparound''  contamination~\cite{Gupta:1993rn,Umeda:2007hy,Feng:2009ij,Dudek:2012gj} before fitting with this formula.

\begin{table*}[t!]
\caption{ \label{tab:MILC_configs}
Simulation parameters of the MILC gauge configurations.
Lattice dimensions are described in lattice units
with spatial ($L$) and temporal ($T$) size.
The gauge coupling  $\beta$ is shown in Column $3$.
The fourth block give bare masses of the light and
strange quark masses in terms of $am_l$ and $am_s$, respectively.
Column $5$ gives pion masses in MeV.
The lattice spatial dimension ($L$) in {\rm fm} and in units of the finite-volume pion mass
are given in Column $6$ and $7$ respectively.
We also list the mass ratio $m_\pi/m_\rho$.
The number of time slices calculated $\pi\pi$ correlators and $\rho$ propagators
for each of the lattice ensembles are shown in Column $9$ and $11$, respectively,
and the last Column gives the number of gauge configurations used in this work.
}
\begin{ruledtabular}
\begin{tabular}{llllcclllll}
Ensemble &$L^3 \times T$ &$\beta$ & $a m_l/a m_s$
& $m_\pi({\rm MeV})$ &$L{\rm(fm)}$  &$m_\pi L$
& $m_\pi/m_\rho$ & $N_{\rm slice}^{\pi\pi}$
& $N_{\rm slice}^{\rho}$  &$N_{\rm cfg}$  \\
\hline
\multicolumn {11}{c}{$a \approx 0.09$~fm}        \\
6496f21b7075m00155m031 & $64^3\times96$  & $7.075$     & $0.00155/0.031$  & $176$
                       & $5.4$           & $4.80$  & $0.224$  & $96$ & $48$  & $60$ \\
4096f21b708m0031m031   & $40^3\times96$  & $7.08$      & $0.0031/0.031$  & $247$
                       & $3.4$           & $4.21$  & $0.297$  & $96$ & $48$  & $400$ \\
4096f3b7045m0031       & $40^3\times96$  & $7.045$     & $0.0031/0.0031$ & $248$
                       & $3.4$           & $4.20$  & $0.303$  & $96$ & $48$  & $400$ \\
3296f21b7085m00465m031 & $32^3\times96$  & $7.085$     & $0.00465/0.031$  & $301$
                       & $2.7$           & $4.11$  & $0.312$  & $96$ & $48$  & $400$ \\
2896f21b709m0062m031   & $28^3\times96$  & $7.09$      & $0.0062/0.031$  & $346$
                       & $2.4$           & $4.14$  & $0.380$  & $96$ & $48$  & $400$ \\
\multicolumn {11}{c}{$a \approx 0.12$~fm}        \\
3264f3b6715m005        & $32^3\times64$  & $6.715$     & $0.005/0.005$  & $275$
                       & $3.7$           & $5.15$  & $0.299$  & $64$ & $32$  & $637$ \\
\end{tabular}
\end{ruledtabular}
\end{table*}

\section{Lattice calculation}
\label{sec:latticeCal}
We employed the MILC gauge configurations with three Asqtad-improved
staggered sea quarks~\cite{Bernard:2010fr,Bazavov:2009bb}.
The simulation parameters are summarized in Table~\ref{tab:MILC_configs}.
By MILC convention, lattice ensembles are referred to as ``coarse''
for the spatial lattice spacing $a\approx0.12$~fm,
and ``fine'' for $a\approx0.09$~fm.
It is handy to adopt $(am_l, am_s)$ to
classify MILC lattice ensembles.
The conjugate gradient method is exploited to calculate the light quark propagators.
We should remember that the MILC gauge configurations are
generated using the staggered formulation of lattice
the fermions~\cite{Kaplan:1992bt} with the fourth root of
fermion determinant~\cite{Bernard:2001av}.
All the gauge configurations were gauge fixed to the Coulomb gauge
before calculating the light quark propagators.

To compute the $\pi\pi$ four-point functions,
the standard conjugate gradient method is adopted to get
the necessary matrix element of the inverse Dirac fermion matrix,
and the periodic boundary condition is applied to
both the spatial and temporal directions.
We compute the correlators on all the time slices,
and explicitly combine the results from all the time slices $T$;
namely, the diagonal correlator $C_{11}(t)$ is measured through
\begin{eqnarray}
 C_{11}(t) &=&
\frac{1}{T}\sum_{t_s=0}^{T} \left\langle
\left(\pi\pi\right)(t+t_s)\left(\pi\pi\right)^\dag(t_s)\right\rangle .
\nonumber
\end{eqnarray}
After averaging the propagators over all the $T$ values,
the statistics are found to be remarkably improved.

For another diagonal correlator $C_{22}(t)$, the $\rho$ correlator,
we calculate
$$
C_{22}(t)=\frac{2}{T}\sum_{t_s=0,2,4,\cdots}^{T}
\left\langle{\rho}^\dag(t+t_s) \rho(t_s)\right\rangle ,
$$
where we sum the correlator over
all the even time slices and average it.

According to the discussion in the Appendix,
the noise-to-signal ratio of the $\rho$ correlator and $\pi\pi$ correlator
are improved as approximately  $\propto \frac{1}{\sqrt{N_{\rm slice} L^3}}$,
where $L$ is the lattice spatial dimension and $N_{\rm slice}$ is the
number of the time slices calculated the propagators for each of the gauge configurations.
In this work, we use the lattice ensembles with relatively large $L$
and sum the $\rho$ correlator over all the even time slices
and the $\pi\pi$ correlator over all the time slices;
consequently, it is natural that the signals of the correlators
should be significantly improved.
Admittedly, the most efficient way to improve the relevant noise-to-signal ratio
is to use finer gauge configurations or anisotropic gauge configurations~\cite{Dudek:2012xn,Wilson:2015dqa}.
See the Appendix for more details.

We evaluate the first off-diagonal correlator $C_{21}(t)$,
the $\pi\pi \to \rho$  three-point function, through
\begin{eqnarray}
C_{21}(t) &=& \frac{1}{T}\sum_{t_s}^{T}
\left\langle \rho(t+t_s)(\pi\pi)^\dag(t_s)\right\rangle ,
\nonumber
\end{eqnarray}
where the summation is  over all the time slice.
Due to the time-reversal symmetry~\cite{Pelissier:2012pi},
we can in practice merely calculate $C_{21}^\ast(t)$.
By the relation $C_{12}(t)=C_{21}^\ast(t)$,
we can freely get the second off-diagonal correlator $C_{12}(t)$.

We measure two-point pion correlators with the zero and nonezero
momenta ($\mathbf{0}$ and $\mathbf{p}$) as well,
\begin{eqnarray}
\label{eq:pi_cor_PW_k000}
C_\pi({\mathbf 0}, t) &=& \frac{1}{T}\sum_{t_s=0}^{T-1}
\langle 0|\pi^\dag ({\mathbf 0}, t+t_s) W_\pi({\mathbf 0}, t_s) |0\rangle, \\
\label{eq:pi_cor_PW_k100}
C_\pi({\mathbf p}, t) &=& \frac{1}{T}\sum_{t_s=0}^{T-1}
\langle 0|\pi^\dag ({\mathbf p}, t+t_s) W_\pi({\mathbf p}, t_s) |0\rangle,
\end{eqnarray}
where $\pi$ is the pion point-source operator
and $W_\pi$ is the pion wall-source operator~\cite{Bernard:2001av,Aubin:2004wf}.
To simplify notation, the summation over the lattice space point in sink
is not written out. It is worth noting that
the summations over all the time slices for $\pi$ propagators
guarantee the extraction of the pion mass with high precision.

Disregarding the contributions from the excited states,
the pion mass $m_\pi$ and energy $E_\pi({\mathbf p})$
can be robustly extracted at large $t$ from the two-point pion correlators~(\ref{eq:pi_cor_PW_k000})
and (\ref{eq:pi_cor_PW_k100}), respectively~\cite{Bazavov:2009bb},
\begin{eqnarray}
\label{eq:pi_fit_PW_k000}
\hspace{-0.6cm} C_\pi({\mathbf 0}, t) &=& A_\pi(\mathbf{0}) \left[e^{-m_\pi t}+e^{-m_\pi(T-t)}\right] +\cdots, \\
\label{eq:pi_fit_PW_k100}
\hspace{-0.6cm} C_\pi({\mathbf p}, t) &=& A_\pi(\mathbf{p})
\left[e^{-E_\pi({\mathbf p}) t}+e^{-E_\pi({\mathbf p})(T-t)}\right] + \cdots,
\end{eqnarray}
where the ellipses show the oscillating parity partners,
and $A_\pi(\mathbf{0})$ and $A_\pi(\mathbf{p})$ are two overlapping amplitudes,
which will be subsequently exploited to estimate the wraparound contributions
for $I=1$ $\pi\pi$ correlators~\cite{Gupta:1993rn,Umeda:2007hy,Feng:2009ij}.

\section{Lattice simulation results }
\label{Sec:Results}

\subsection{Pion mass and dispersion relation}
\label{Sec:Results_pionMass}
\begin{figure}[tp!]
\includegraphics[width=8.5cm,clip]{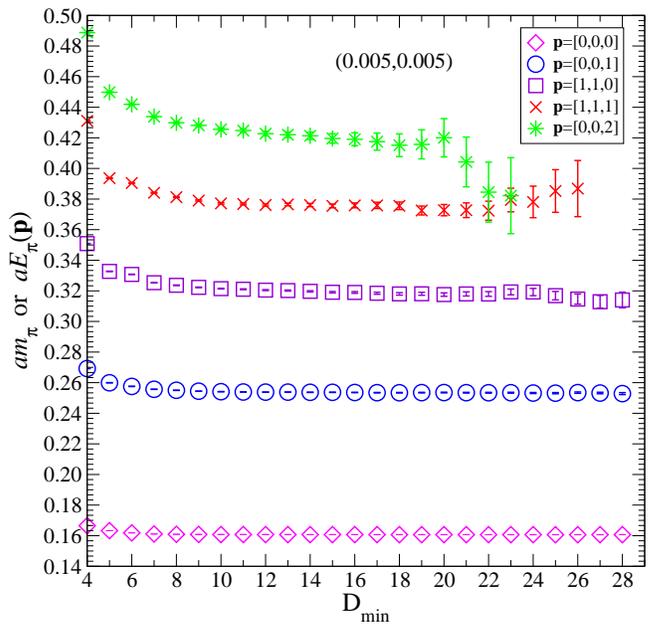}
\caption{
\label{fig:eff_pion_6715}
(color online). Effective pion mass $m_\pi$ or energy $E_{\pi}(\mathbf{p})$
plots as the functions of $\rm D_{min}$
for the $(0.005,0.005)$ ensemble.
The plateaus are quickly reached  typically at or before $t=8$ from the source.
}
\end{figure}

For each of the lattice ensembles, the pion masses $m_\pi$ and
energies $E_{\pi}(\mathbf{p})$ were cautiously selected
by seeking a combination of a plateau
in the mass (or energy) as a function of the minimum fitting distances
a $\rm D_{min}$~\cite{Bernard:2001av,Aubin:2004wf},
fit quality, and $\rm D_{min}$ large enough to efficiently suppress the excited states~\cite{Feng:2009ij,Aubin:2004fs}.
For example, Fig.~\ref{fig:eff_pion_6715} exhibits the fit results
of the pion masses or pion energies in lattice units as a function of $\rm D_{min}$ for
the $(0.005,0.005)$ ensemble.
It is interesting and important to note that
the rapid relaxations to the ground state for all of the five momenta,
typically at or before $t=9$ from the source,
indicates the feasibility of the wall-source and point-sink pion interpolators.

The lattice-measured values of the pion masses $m_\pi$ and pion energies $E_{\pi}(\mathbf{p})$
in lattice units, along with the fit ranges and fit qualities,
are tabulated in Table~\ref{tab:m_pi_pi}.
The overlapping amplitudes $A_\pi(\mathbf{0})$ or $A_\pi(\mathbf{p})$
denoted in Eqs.~(\ref{eq:pi_fit_PW_k000}) and~(\ref{eq:pi_fit_PW_k100})
are also listed in Table~\ref{tab:m_pi_pi};
these are later used to estimate the wraparound pollution to the $I=1$ $\pi\pi$ four-point correlators~\cite{Gupta:1993rn,Umeda:2007hy,Feng:2009ij}.
Note that the ETMC Collaboration reduces this unwanted lattice wraparound artifact
by choosing the maximum time of the fit range to be far enough
from the temporal boundaries~\cite{Feng:2010es}.
In the present work, our measured quantities from these two-point functions
are sufficiently precise to allow us to subtract the wraparound contributions.

\begin{table*}[t]	
\caption{\label{tab:m_pi_pi}
Summary of the pion masses $a m_\pi$ or pion energies $a E_\pi(\mathbf{p})$
obtained from the pion propagators for six MILC lattice ensembles with four momenta,
$\mathbf{p} =[0,0,1]$, $[1,1,0]$, $[1,1,1]$, and $[0,0,2]$.
The lattice-measured pion energies $a E_\pi(\mathbf{p})$ are compared with
the analytical predictions from the continuum (\ref{eq:mass_disp})
and free lattice theory (\ref{eq:mass_lat_dr}),
where the uncertainties are estimated solely from the statistical errors of the lattice-measured $a m_{\pi}$.
The fifth block shows the overlapping amplitudes $A_\pi(\mathbf{0})$ or $A_\pi(\mathbf{p})$ denoted
in Eqs.~(\ref{eq:pi_fit_PW_k000}) and (\ref{eq:pi_fit_PW_k100}), respectively.
The third and fourth blocks indicate the fit range, and fit quality $\chi^2/{\rm DOF}$, respectively.
}
\begin{ruledtabular}
\begin{tabular}{ccccclcc}
$\rm Ensemble$ & $\mathbf{n}=\mathbf{p} \tfrac{L}{2\pi}$
& Range & $\chi^2$/{\rm DOF} & $A_\pi(\mathbf{0})$/$A_\pi(\mathbf{p})$
& $a m_\pi/a E_\pi(\mathbf{p})$ & $a E_{cont}$ & $a E_{lat}$   \\
\hline
\multirow{6}*{$(0.00155, 0.031)$}
& $(0,0,0)$      & $33-48$       & $14.1/12$     & $6603.86\pm4.84$
& $0.07501(6)$   & $-$           & $-$           \\
& $(0,0,1)$      & $19-45$       & $26.1/21$     & $4136.59\pm7.49$
& $0.12348(14)$  & $0.12355(6)$  & $0.12345(6)$  \\
& $(1,1,0)$      & $16-42$       & $18.4/23$     & $3035.31\pm12.73$
& $0.15793(19)$  & $0.15780(5)$  & $0.15760(4)$  \\
& $(1,1,1)$      & $15-40$       & $26.2/22$     & $2324.42\pm16.12$
& $0.18579(24)$  & $0.18585(4)$  & $0.18553(4)$  \\
& $(0,0,2)$      & $13-36$       & $23.6/20$     & $2055.65\pm18.74$
& $0.21023(36)$  & $0.21019(4)$  & $0.20951(4)$  \\
\hline
\multirow{6}*{$(0.0031, 0.0031)$}
& $(0,0,0)$      & $30-47$       & $15.8/14$     & $1091.62\pm1.73$
& $0.10505(6)$   & $-$           &  $-$          \\
& $(0,0,1)$      & $19-48$       & $23.9/26$     & $619.13\pm2.37$
& $0.18963(38)$  & $0.18896(5)$  & $0.18857(5)$  \\
& $(1,1,0)$      & $16-48$       & $37.8/29$     & $482.97\pm3.24$
& $0.24709(95)$  & $0.24572(4)$  & $0.24493(4)$  \\
& $(1,1,1)$      & $13-36$       & $29.2/20$     & $414.45\pm7.44$
& $0.2929(15)$   & $0.29164(3)$  & $0.29038(3)$  \\
& $(0,0,2)$      & $12-48$       & $38.3/33$     & $381.03\pm7.93$
& $0.3343(18)$   & $0.33125(3)$  & $0.32856(3)$  \\
\hline
\multirow{6}*{$(0.0031, 0.031)$}
& $(0,0,0)$      & $33-45$       & $11.1/9$      & $1218.53\pm1.74$
& $0.10535(6)$   & $-$           & $-$           \\
& $(0,0,1)$      & $19-41$       & $24.1/19$     & $684.89\pm1.89$
& $0.19016(21)$  & $0.18916(6)$  & $0.18875(6)$  \\
& $(1,1,0)$      & $16-36$       & $16.4/17$     & $525.61\pm3.83$
& $0.24713(49)$  & $0.24588(5)$  & $0.24506(4)$  \\
& $(1,1,1)$      & $15-37$       & $16.4/19$     & $431.92\pm6.32$
& $0.29124(105)$ & $0.29177(4)$  & $0.29049(4)$  \\
& $(0,0,2)$      & $13-34$       & $22.6/18$     & $398.53\pm9.34$
& $0.33244(172)$ & $0.33135(4)$  & $0.32866(3)$  \\
\hline
\multirow{6}*{$(0.00465, 0.031)$}
& $(0,0,0)$      & $30-48$       & $10.3/15$     & $538.41\pm1.31$
& $0.12852(9)$   & $-$           &  $-$            \\
& $(0,0,1)$      & $20-45$       & $30.9/22$     & $292.55\pm2.32$
& $0.23513(46)$  & $0.23465(10)$ & $0.23390(10)$   \\
& $(1,1,0)$      & $17-43$       & $28.2/23$     & $224.28\pm5.03$
& $0.3048(13)$   & $0.30596(8)$  & $0.30442(8)$    \\
& $(1,1,1)$      & $15-27$       & $14.2/9$      & $197.07\pm6.46$
& $0.3643(23)$   & $0.36355(6)$  & $0.36110(6)$    \\
& $(0,0,2)$      & $13-24$       & $10.6/8$      & $172.67\pm9.00$
& $0.4132(42)$   & $0.41318(6)$  & $0.40798(6)$   \\
\hline
\multirow{5}*{$(0.0062, 0.031)$}
& $(0,0,0)$      & $30-48$       & $24.3/15$    & $330.73\pm0.94$
& $0.14718(13)$  & $-$           &  $-$           \\
& $(0,0,1)$      & $20-48$       & $34.3/25$    & $179.83\pm2.31$
& $0.26751(72)$  & $0.26837(14)$ & $0.26725(14)$  \\
& $(1,1,0)$      & $17-42$       & $22.7/22$    & $139.60\pm3.65$
& $0.3471(17)$   & $0.34982(11)$ & $0.34752(11)$   \\
& $(1,1,1)$      & $14-19$       & $3.0/2$      & $126.23\pm4.27$
& $0.4170(26)$   & $0.41561(9)$  & $0.411987(9)$   \\
\hline
\multirow{6}*{$(0.005, 0.005)$}
& $(0,0,0)$      & $20-32$       & $6.8/9$       & $1717.95\pm1.66$
& $0.16068(5)$   & $-$           & $-$           \\
& $(0,0,1)$      & $14-32$       & $22.4/15$     & $1091.01\pm1.96$
& $0.25371(13)$  & $0.25373(3)$  & $0.25292(3)$  \\
& $(1,1,0)$      & $13-30$       & $23.4/13$     & $856.21\pm3.86$
& $0.32013(35)$  & $0.32083(3)$  & $0.31917(3)$  \\
& $(1,1,1)$      & $12-32$       & $11.7/17$     & $737.78\pm6.11$
& $0.37610(71)$  & $0.37614(2)$  & $0.37355(2)$  \\
& $(0,0,2)$      & $11-32$       & $22.6/18$     & $669.81\pm8.00$
& $0.42475(112)$ & $0.42430(2)$  & $0.41897(2)$
\end{tabular}
\end{ruledtabular}
\end{table*}

The rho masses $m_{\rho}$ are extracted from the $\rho$  correlator,
and the mass ratios of $m_\pi/m_\rho$ are listed in Table~\ref{tab:MILC_configs}.
It is important to note that our lattice-measured pion masses and
the mass ratios of $m_\pi/m_\rho$
turn out to be in good agreement with the corresponding MILC
determinations~\cite{Bernard:2007ps,Bernard:2001av,Aubin:2004wf,Bazavov:2009bb}.
Note that our simulations are all carried out at physical kinematics
$m_\pi/m_\rho < 1/2$.

It is interesting and important to note that the pion mesons on the lattice
are found to have a continuumlike dispersion relation,
as already observed in Ref.~\cite{Lang:2012sv}.
Say more specifically, for $\mathbf{n}^2<4$,
our lattice-measured pion energies $E_\pi(\mathbf{p} = \tfrac{2\pi}{L}\mathbf{n})$
are in good keeping with the continuum dispersion relation
\begin{equation}
\label{eq:mass_disp}
E_{\rm cont}({\mathbf p}) = \sqrt{ m_\pi^2 + |{\mathbf p}|^{2} }
\end{equation}
within the errors, and are better than
those evaluated with the prediction of the free lattice theory
\begin{equation}
\label{eq:mass_lat_dr}
aE_{\rm lat} = \cosh^{-1}\Big[\cosh(am_\pi) + 2\sum_i \sin^2\left(\tfrac{1}{2}a p_i\right)\Big],
\end{equation}
where $p=|{\mathbf p}|$ is the magnitude of each pion's scattering momentum
in the center-of-mass frame.
This is probably because the rotational invariance properties are improved
due to the hypercubic smearing of the gauge link
and quark operators~\cite{Lang:2012sv}.\footnote{
This is also probably due to the significant improvement of the signal of the pion propagator
as the pion momentum increases with large-enough $L$~\cite{DellaMorte:2012xc,Parisi:1983ae}.
As previously explained,
the noise-to-signal ratio $R_{NS}(t)$ of the pion energy
$E_\pi(\mathbf{p} = \tfrac{2\pi}{L}\mathbf{n})$
usually grows exponentially as
$
R_{NS}(t) \propto \frac{1}{\sqrt{L^3}}  \exp {\left(\sqrt{m_{\pi}^2 + \tfrac{4\pi^2}{L^2}\mathbf{n}^2}-m_\pi\right)t}.
$
}

Note that for $\mathbf{n}^2=4$ (i.e., $\mathbf{p}=[0,0,2]$),
our lattice-measured pion energies $E_\pi(\mathbf{p})$
for the $(0.0031,0.031)$ $[L=40]$, $(0.0031,0.0031)$ $[L=40]$, $(0.00465,0.031)$ $[L=32]$
and $(0.005,0.005)$ $[L=32]$ ensembles
are well consistent with the continuum dispersion relation (\ref{eq:mass_disp}),
whilst, those of the $(0.0062,0.031) [L=28]$ ensemble only barely
meet the continuum dispersion relation (\ref{eq:mass_disp}).
For this reason, we ignore the calculations relevant to
the momentum  $\mathbf{p}=[0,0,2]$ for the $(0.0062,0.031)$ ensemble.

In summary, the continuum dispersion relation (\ref{eq:mass_disp}) for the single pion state
is valid up to the momentum $\mathbf{p}=[0,0,2]$
within statistical errors [except for the $(0.0062,0.031)$ ensemble, the relevant results for which are not listed in Table~\ref{tab:m_pi_pi}].
Consequently, we will use the continuum dispersion relation (\ref{eq:mass_disp})
throughout the remaining analysis.
This means that the center-of-mass scattering momentum $p^*$ is
extracted from the lattice energy through the continuum dispersion relation,
and the resulting $p^*$
is used to extract the scattering phase shift.
Additional relevant issues will be discussed in Sec.~\ref{SubSec: Effect of finite lattice spacing}.
It is worth stressing that the robust measurements of the pion propagator with high momenta
indeed guarantee reliable estimations of the $\pi\pi$ propagator with high momenta.

\subsection{Finite-$T$ contributions for the $\pi\pi$ correlators}
\label{sec:Lat_artifact}
In this work, $I=1$ $\pi\pi$ energy spectra are meticulously secured
from  $\pi\pi$ correlators,
which are unavoidably impacted by the finite temporal extent of the lattice~\cite{Gupta:1993rn,Umeda:2007hy,Feng:2009ij,Dudek:2012gj}.
In principle, the size of the finite temporal effects can be estimated,
and are slight on a typical lattice study.
Nonetheless, these  effects are large enough to be visible,
particularly for the $I=1$ $\pi\pi$ correlators calculating
with small quark masses~\cite{Fu:2011bz}.
Using the original notations in~\cite{Dudek:2012gj}, we here briefly review the finite-$T$ effects
on $I=1$ $\pi\pi$ correlators at rest (i.e., the total momentum of $\pi\pi$ system $\mathbf{P} = \mathbf{0}$)
and those in flight (i.e., $\mathbf{P} \neq \mathbf{0}$).

Since the periodic boundary condition is enforced in the temporal direction,
one of two pions can spread $T-t$ time steps backwards,
which leads to a pollution of the $\pi\pi$ correlators at large $t$~\cite{Gupta:1993rn,Umeda:2007hy,Feng:2009ij}.
Additionally, in the isospin limit,
two direct quark-line diagrams ($D$ and $X$) in Fig.~\ref{fig:diagram}
contribute to the $I=1$ $\pi\pi$ scattering amplitudes,
and both of them have wraparound pollution.
The wraparound pollution from the direct diagram $D$
in the limit of weakly interacting pions,
which is one pion with the momentum ${\mathbf p}$,
and another pion with momentum ${\mathbf q}$,
can be approximately estimated by~\cite{Dudek:2012gj}
\begin{eqnarray}
{\rm WP}(t) &\approx&  A_\pi(\mathbf{q}) A_\pi(\mathbf{p})
\left( e^{-E_\pi(\mathbf{q}) (T-t)}
e^{-E_\pi(\mathbf{p}) t} \right.\cr
 &&\left. +  e^{-E_\pi(\mathbf{p}) (T-t)} e^{-E_\pi(\mathbf{q}) t} \right),
\label{D_wrap}
\end{eqnarray}
where the overlapping amplitudes $A_\pi(\mathbf{p})$
denoted in~(\ref{eq:pi_fit_PW_k100}),
along with the pion masses $m_\pi$ and pion energy $E_\pi(\mathbf{p})$,
can be robustly extracted from the pion propagators.
The undesired wraparound contributions to
the $I=1$ $\pi\pi$ four-point correlators in the moving frame are generally time dependent~\cite{Dudek:2012gj}.
The wraparound pollution of the direct diagram $X$ in Fig.~\ref{fig:diagram}
can be analogously dealt with;
therefore, we do not explicitly write it out.

As a simple example, considering the $I=1$ $\pi\pi $ correlators
in  CMF (${\mathbf{q}} = -{\mathbf{p}}$),
this leads to a constant pollution 
\begin{equation}
\label{eq:D_CM_fake_diagram}
C(\mathbf{p}) = 2(A_\pi(\mathbf{p}))^2 e^{-E_\pi(\mathbf{p}) T} ,
\end{equation}
where the overlapping amplitudes $A_\pi(\mathbf{p})$, and pion energy $E_\pi(\mathbf{p})$
are summarized in Table~\ref{tab:m_pi_pi}.

Considering another concrete example of the $\pi\pi$ correlator
with $\pi^+(\mathbf{0}) \pi^+(\mathbf{p})$ at the source and
     $\pi^-(\mathbf{0}) \pi^-(\mathbf{p})$ at the sink
     (trhis is one pion at rest, one pion with the momentum ${\mathbf p}$,
and total momentum  ${\mathbf P} = {\mathbf p}$),
the wanted contribution in the limit of weakly interacting pions
can be approximately estimated by~\cite{Dudek:2012gj}
\begin{equation}
\approx A_\pi(\mathbf{0}) A_\pi(\mathbf{p})
e^{-(m_\pi + E_\pi(\mathbf{p}) ) t},
\label{flight_good}
\end{equation}
where the overlapping amplitudes $A_\pi(\mathbf{0})$ and  $A_\pi(\mathbf{p})$
are denoted in Eqs.~(\ref{eq:pi_fit_PW_k000})
               and (\ref{eq:pi_fit_PW_k100}), respectively.
Meanwhile, the wraparound terms for this moving frame ${\rm WP}_{\rm MF}(t)$
are evaluated by~\cite{Dudek:2012gj}
\begin{eqnarray}
{\rm WP}_{\rm MF}(\mathbf{p}, t) &\approx&  A_\pi(\mathbf{0}) A_\pi(\mathbf{p})
\left( e^{-m_\pi (T-t) } e^{-E_\pi(\mathbf{p}) t} \right.\cr
 &&\left. +  e^{-E_\pi(\mathbf{p}) (T-t) } e^{-m_\pi t} \right),
\label{flight_wrap}
\end{eqnarray}
where the first term is anticipated to
lead the contamination for the time regions of interest~\cite{Dudek:2012gj}.
Moreover, the largest pollution term is not
a constant but rather has a time dependence $\sim e^{- \Delta E_\pi t}$,
where $\Delta E_\pi \equiv E_\pi(\mathbf{p}) - m_\pi$ is  the positive energy gap
between one pion with zero momentum
and another with momentum ${\mathbf p}$.
Besides, the second pollution term has a time dependence $\sim e^{\Delta E_\pi t}$.
Note that the ratio of the first term to the second term is roughly proportional to
$e^{-\Delta E_\pi (T-2t)}$, which indicates that both pollution terms significantly
contribute the whole pollution on large times.

Since the impact of the finite-$T$ effects on the $I=1$ $\pi\pi$ correlators in flight
is statistically significant~\cite{Dudek:2012gj},
it is necessary to correct these terms in the variational analysis of the in-flight $\pi\pi$ spectra.
In practice, we subtract all the pollution terms
in Eq.~(\ref{flight_wrap}) from the $\pi\pi$ correlators.
This turns out to be a rather good approximation
for the lattice simulation in this work.

\begin{figure}[t]
\includegraphics[width=8.0cm,clip]{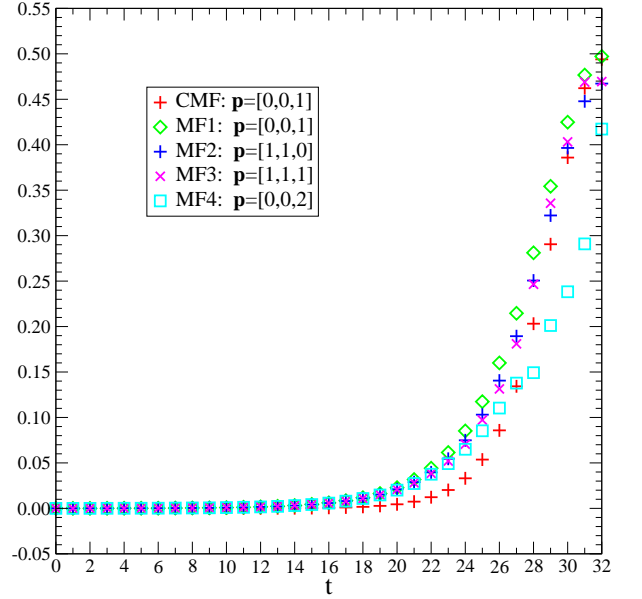}
\caption{\label{fig:ratio_WC}
(color online). Ratios of the wraparound pollution
to the $\pi\pi$ correlators of the direct diagram for the $(0.005,0.005)$ ensemble
using Eq.~(\ref{eq:ratio_WC}) for four momenta,
$\mathbf{p} =[0,0,1]$, $[1,1,0]$, $[1,1,1]$, and $[0,0,2]$,
along with the ratio for the center-of-mass frame at $\mathbf{p} =[0,0,1]$.
These ratios are generally near  $1/2$  as $t$ approaches
$T/2$ as anticipated from the analytical statements in Refs.~\cite{Gupta:1993rn,Umeda:2007hy,Feng:2009ij,Dudek:2012gj}.
}
\end{figure}

In order to comprehend this finite-$T$ effect at a quantitative level,
we denote a quantity
\begin{equation}
\label{eq:ratio_WC}
R_{\rm MF}(\mathbf{p}, t) =
\frac{ {\rm WP}_{\rm MF}(\mathbf{p},t)}{D_{\pi\pi}^{I=1}(\mathbf{p},t)},
\end{equation}
which is the ratio of the finite-$T$ effect ${\rm WP}_{\rm MF}(\mathbf{p},t)$
calculated by Eq.~(\ref{flight_wrap})
to the $I=1$ $\pi\pi$ correlator $D_{\pi\pi}^{I=1}(\mathbf{p},t)$  of the direct diagram $D$.
In Fig.~\ref{fig:ratio_WC}, we illustrate this ratio for the ($0.005, 0.005$) ensemble at four momenta,
$\mathbf{p}$ $=[0,0,1]$, $[1,1,0]$, $[1,1,1]$, and $[0,0,2]$,
together with the ratio for the center-of-mass frame,
which is denoted by
$$R_{\rm CM}(\mathbf{p},t) =
C(\mathbf{p})/D_{\pi\pi}^{I=1}(\mathbf{p},t) $$
for $\mathbf{p} =[0,0,1]$,
where $C(\mathbf{p})$ is defined in Eq.~(\ref{eq:D_CM_fake_diagram}).
These ratios turn out to make a significant contribution to $\pi\pi$ correlators
as $t$ approaches $T/2$~\cite{Gupta:1993rn,Umeda:2007hy,Feng:2009ij,Dudek:2012gj}.
Consequently, it is required to explicitly
account for this pollution when extracting the $\pi\pi$ energy.
Through appropriately subtracting this effect from the $\pi\pi$ correlators,
these unwanted finite-$T$ effects are anticipated to  be neatly removed.
The relevant ratios for the $(0.005,0.005)$ ensemble are illustrated
in Fig.~\ref{fig:LM_6715}.
It is interesting to note that the wraparound pollution generally
contributes in relatively smaller quantities for the higher momenta.

\subsection{Energy eigenvalues}
The finite-$T$ effects for $\pi\pi$  correlators at rest
are constant in time, while those in flight are generally time-dependent.
It is natural to explicitly incorporate these
wraparound terms for a successful energy spectral fit
of  $\pi\pi$ correlators.
In the present study, these wraparound effects can be accurately estimated and consequently, can be appropriately
subtracted from the corresponding $I=1$ $\pi\pi$ correlators.
After deducting these undesired effects, the remaining $I=1$ $\pi\pi$ correlators
then hold clean information.

\begin{figure}[thp!]
\includegraphics[width=8.2cm]{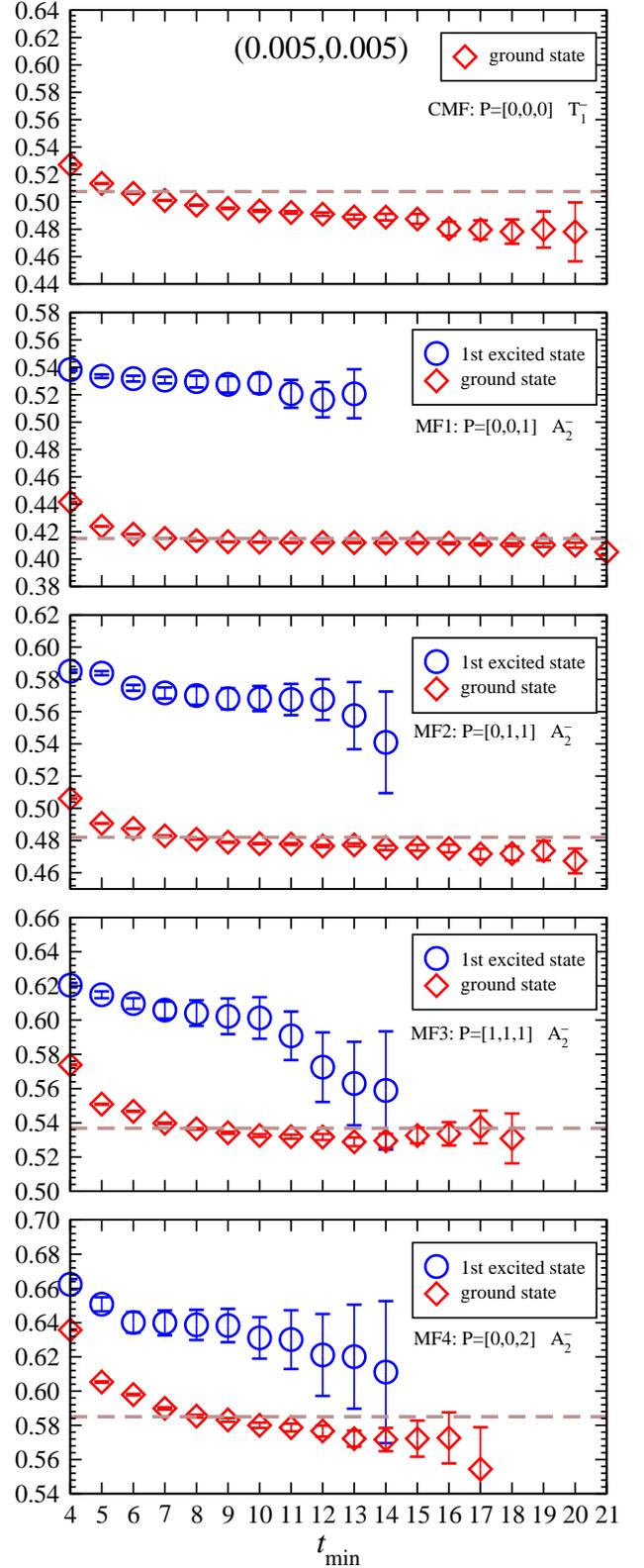}
\caption{
Fitted energy levels as functions of $\rm {\it t}_{min}$ for the ground states of the ${\mathbf T}_1^-$ representation in CMF,
and for the ground and first excited states of the ${\mathbf A}_2^-$, ${\mathbf B}_1^-$,
${\mathbf A}_2^-$ and ${\mathbf A}_2^-$ representations for MF1, MF2, MF3, and MF4, respectively,
with the $(0.005,0.005)$ ensemble.
}
\label{fig:LM_6715}
\end{figure}

As practised in Refs.~\cite{Fu:2012gf},
we calculate two eigenvalues $\lambda_n(t,t_R)$ ($n=1,2$) for the matrix $M(t,t_R)$
denoted in Eq.~(\ref{eq:M_def}) with the reference time $t_R$.
By defining a fit range $[t_{\rm{min}}, t_{\rm{max}}]$
and adjusting the minimum fitting distance $t_{\rm{min}}$
and maximum fitting distance $t_{\rm{max}}$,
we can acquire energy levels from $\lambda_n(n=1,2)$ in a correct manner.
In this work, we take $t_\mathrm{min}=t_R+1$~\cite{Feng:2010es};
in order to extract the desired energies $\overline{E}_n (t_\mathrm{min})$ $(n=1,2)$, two eigenvalues $\lambda_n(t,t_R)(n=1,2)$ at the chosen $t_\mathrm{min}$
were fit to Eq.~(\ref{Eq:asy}), with the $t_{\rm{max}}$ either at $T/2$
or where the fractional statistical errors exceeded about $20\%$
for two successive time slices.
Examples of fitted energy levels as functions of $t_{\rm{min}}$
for the nine energy states with the relevant representations considered in the present work
are illustrated in Fig.~\ref{fig:LM_6715} for the $(0.005,0.005)$ ensemble.
The dotted lines in each panel
indicate the energy levels of the two free pions for the relevant representation.

For each of the lattice ensembles, the energy levels $\overline{E}_n (n=1,2)$ with
the ${\mathbf A}_2^-$, ${\mathbf B}_1^-$, ${\mathbf A}_2^-$, and ${\mathbf A}_2^-$ representations for the MF1, MF2, MF3, and MF4, respectively,
were carefully selected by seeking the combination of
a plateau in the effective energy plots
as the function of $t_{\rm{min}}$ and a reasonable fit quality.
The fit range and fit quality $\chi^2/\mathrm{DOF}$, along with the fitted $\overline{E}_n$ ($n=1,2$)
for six MILC lattice ensembles,
are summarized in Table~\ref{tab:fitting_results}.
The lattice-measured energy levels $\overline{E}_n$ $(n=1,2)$
are then employed to derive the $p$-wave scattering phase shifts $\delta_1$ by the corresponding finite size formulas,
which are summarized in Table~\ref{tab:fitting_results}.
The relevant fitted results with
the ${\mathbf T}_1^-$  representation for CMF
are also summarized in Table~\ref{tab:fitting_results} as well.\footnote{
The lattice determinations of the four-pion thresholds for the  MFs and CMFs
are generally discussed in Ref.~\cite{Bali:2015gji}.
Moreover, according to the discussions in Refs.~\cite{Bali:2015gji,Akhmetshin:1999ty},
the $\rho$ meson to $4\pi$ states is indeed negligible.
}

It is worthwhile stressing that the finite-size effects are exponentially suppressed with the combination $m_\pi L$,
which obviously decreases with the small $a m_\pi$; it is expensive to compensate for this with higher lattice spatial dimensions $L$.
From Table~\ref{tab:MILC_configs}, we note that in the present study our lattice volumes all have $m_\pi L > 4$;
consequently, the finite-size effects are negligible,
and the L\"uscher formulas are perfectly satisfied~\cite{Luscher:1990ux,Luscher:1990ck}.

\newpage
\LTcapwidth=496pt
\begin{longtable*}[htp!]{@{\extracolsep{\fill}}ccccccccc@{}}
\caption[fitting_results]{
Summaries of the fitted energy levels with the ${\mathbf A}_2^-$ representation for the ground state ($n=1$) and first excited state ($n=2$)
in MF1, MF2, MF3, and MF4, respectively,
and the ${\mathbf T}_1^-$  representation in CMF, for six lattice ensembles.
The fit range ($t_{\mathrm{min}}, t_{\mathrm{max}}$),
$\chi^2/\mathrm{DOF}$,
and extracted energy levels $E_n$ ($n=1,2$) are tabulated in Column 4, 5, 6, respectively.
The center-of-mass scattering momentum $p^*$ and the invariant mass  $\sqrt{s}$ are
obtained using the dispersion relations~(\ref{eq:Disp_Two_Cont_k}),
and the relevant $p$-wave scattering phase $\delta_1$ in units of degree
is obtained by the corresponding finite-size formulas.
}
\label{tab:fitting_results}\\

\hline \hline \\[0.1ex]
\multicolumn{1}{c}{${\rm Ensemble}$}  & \multicolumn{1}{c}{Frame} & \multicolumn{1}{c}{Level $n$}
& \multicolumn{1}{c}{Fit range} & \multicolumn{1}{c}{$\chi^2/\mathrm{DOF}$} &
\multicolumn{1}{c}{$aE_n$} & \multicolumn{1}{c}{$a\sqrt{s}$ } & \multicolumn{1}{c}{$ap^*$} & \multicolumn{1}{c}{$\delta_1(^\circ)$} \\[0.5ex] \hline \\

\linespread{5pt}
\\[0.1ex]
\multirow{9}*{$(0.00155,0.031)$}
& CMF & & $21-40$ & 21.5/16 & 0.2446(17) &         & 0.00933(21) & 6.0(3.9)  \\ \\[0.25ex]
& \multirow{2}*{MF1}
 &1&$22-44$ & 15.1/19 & 0.19779(21) & 0.17170(24) & 0.00174(2)  & 0.69(20)  \\
&&2&$9-34$  & 26.4/22 & 0.3574(18)  & 0.3437(19)  & 0.02390(32) & 100.4(8.1) \\ \\[0.10ex]
& \multirow{2}*{MF2}
 &1&$16-38$ & 21.3/19 & 0.23190(32) & 0.18575(40) & 0.00300(4)  & 1.28(53) \\
&&2&$8-20$ & 9.6/9   & 0.3758(16)  & 0.3492(17)  & 0.02486(30) & 103.5(6.9) \\ \\[0.10ex]
& \multirow{2}*{MF3}
 &1&$14-24$ & 9.1/7 & 0.25950(48) & 0.19602(64) & 0.00398(6)  & 1.86(1.6) \\
&&2&$8-18$  & 8.2/7 & 0.3988(19)  & 0.3607(21)  & 0.02691(38) & 132.8(4.1) \\ \\[0.10ex]
& \multirow{2}*{MF4}
 &1&$13-22$ & 8.2/6  & 0.28404(71) & 0.20525(98) &  0.00490(10) & 2.5(2) \\
&&2&$8-18$  & 7.2/7 & 0.4153(21)  & 0.3659(24)  &  0.02785(44) & 131.59(5.3) \\ \\[0.05ex]
\hline \\[0.1ex]
\multirow{9}*{$(0.0031,0.0031)$}
& CMF & & $22-48$ & 28.2/23 & 0.3355(64) && 0.0171(11) & 53.6(7.2)  \\ \\[0.10ex]
& \multirow{2}*{MF1}
 &1&$18-37$ & 15.6/16 & 0.29003(40) & 0.24381(48) & 0.00383(6)  & 1.87(19)  \\
&&2&$9-35$  & 20.7/23 & 0.3841(21)  & 0.3505(23)  & 0.01968(40) & 96.8(2.8) \\ \\[0.10ex]
&  \multirow{2}*{MF2}
 &1&$15-40$ & 20.8/22 & 0.34589(81) & 0.26512(106)& 0.00654(14) & 3.81(75)  \\
&&2&$8-28$  & 19.6/17 & 0.4214(23)  & 0.3581(27)  & 0.02103(48) & 109.4(3.2) \\ \\[0.10ex]
&  \multirow{2}*{MF3}
 &1&$14-39$ & 27.4/22 & 0.38929(133)& 0.2784(19) &  0.00835(26)  & 5.8(1.4) \\
&&2&$8-22$  & 13.2/11 & 0.4532(27)  & 0.3625(34) &  0.02181(61)  & 123.9(4.6) \\ \\[0.10ex]
&  \multirow{2}*{MF4}
 &1&$12-38$ & 13.7/23 & 0.4283(25)  & 0.2911(36) &  0.01016(52)  & 9.7(4.2) \\
&&2&$8-18$  & 11.2/7  & 0.4807(36)  & 0.3639(48) &  0.02207(87)  & 123.6(5.9) \\ \\[0.1ex]
\hline \\[0.1ex]
\multirow{9}*{$(0.0031,0.031)$}
& CMF & & $21-32$ & 15.1/8 & 0.3423(61) &        &  0.0182(10)  & 46.3(7.0)  \\ \\[0.25ex]
& \multirow{2}*{MF1}
 &1&$22-41$ & 18.4/16 & 0.29134(62) & 0.24543(75) & 0.00395(9) & 1.51(31)  \\
&&2&$9-32$ & 19.5/20 & 0.3900(18)  & 0.3570(20)  & 0.0208(35) & 89.4(2.5) \\ \\[0.10ex]
& \multirow{2}*{MF2}
 &1&$16-36$ & 20.6/17 & 0.34632(99) & 0.2657(13)  & 0.00655(17) & 3.8(0.9)  \\
&&2&$8-20$  & 8.8/9   & 0.4230(27)  & 0.3600(32)  & 0.02130(57) & 107.8(3.8) \\ \\[0.10ex]
& \multirow{2}*{MF3}
 &1&$14-22$ & 8.1/5   & 0.3899(15)  & 0.2786(21) & 0.00830(29)  & 6.0(1.6) \\
&&2&$8-18$  & 8.0/7   & 0.4590(24)  & 0.3697(30) & 0.02307(55)  & 116.3(4.3) \\ \\[0.10ex]
& \multirow{2}*{MF4}
 &1&$13-21$ & 9.3/5  & 0.4288(17) & 0.2918(25) &  0.01019(36)  & 9.6(2.8)   \\
&&2&$7-17$  & 10.1/7 & 0.4873(32) & 0.3725(42) &  0.02359(77)  & 115.0(5.3) \\ \\[0.05ex]
\hline \\[0.05ex]
\multirow{9}*{$(0.00465,0.031)$}
&CMF&&$10-18$&7.2/5 & 0.3769(61) &            & 0.0190(15)  & 86.7(6.7) \\ \\[0.10ex]
& \multirow{2}*{MF1}
 &1&$15-35$&28.6/17 & 0.3539(14) & 0.2944(17) & 0.00516(25) & 3.09(40)   \\
&&2&$7-28$ &24.9/18 & 0.4381(26) & 0.3916(29) & 0.0218(6)   & 131.2(2.2) \\ \\[0.10ex]
& \multirow{2}*{MF2}
 &1&$16-48$&34.3/29 & 0.4214(42) & 0.3170(56) & 0.00862(88) & 7.8(2.9)   \\
&&2&$7-27 $&29.8/17 & 0.4849(34) & 0.3975(41) & 0.0230(8)   & 144.2(3.1) \\ \\[0.10ex]
& \multirow{2}*{MF3}
 &1&$15-36$&22.1/18 & 0.4741(63) & 0.3303(90) & 0.0108(15)  & 11.5(6.2)  \\
&&2&$8-21$ &15.9/10 & 0.5203(44) & 0.4070(57) & 0.0249(12)  & 154.7(4.2) \\ \\[0.10ex]
& \multirow{2}*{MF4}
 &1&$12-28$&19.8/13 & 0.5219(56) & 0.3438(86) &  0.0130(15) & 18.7(7.8)  \\
&&2&$8-20$ &14.7/9  & 0.5705(52) & 0.4138(73) &  0.0263(15) & 152.0(6.6) \\ \\[0.05ex]
\hline \\[0.1ex]
\multirow{7}*{$(0.0062,0.031)$}
& CMF & &$17-48$ & 34.5/28 & 0.397(11)  &  &  0.0177(23) & 110.7(4.0) \\ \\[0.10ex]
& \multirow{2}*{MF1}
 &1&$12-17$&1.3/2  &0.4018(12)&0.3333(13) & 0.00611(23) & 3.69(21)   \\
&&2&$7-20$ &13.0/10&0.4716(25)&0.4148(28) & 0.02135(59) & 150.6(1.7) \\ \\[0.10ex]
& \multirow{2}*{MF2}
 &1&$12-32$&21.0/16&0.4766(40)&0.3559(51)&0.01001(91)& 10.3(2.2)  \\
&&2&$7-17/$&9.8/7  &0.5367(48)&0.4328(60)&0.0252(13) & 156.3(3.6) \\ \\[0.10ex]
& \multirow{2}*{MF3}
 &1&$13-21$ & 5.9/5  & 0.5304(66) & 0.3608(98) & 0.0109(18) & 20.2(8.0)  \\
&&2&$6-16/$ & 8.4/7  & 0.5923(67) & 0.4470(89) & 0.0283(20) & 163.7(5.0) \\ \\[0.05ex]
\hline \\[0.1ex]
\multirow{9}*{$(0.005,0.005)$}
& CMF  &  & $18-24$ & 2.9/3 & 0.4715(89)&  &  0.0298(21) & 40.4(9.2) \\ \\[0.10ex]
& \multirow{2}*{MF1}
 &1&$11-18$ & 2.7/4 & 0.41110(25)) & 0.36118(28) &  0.00680(5) & 1.49(12) \\
&&2&$8-18$  & 8.2/7 & 0.5296(43)   & 0.4919(46)  & 0.0347(11)  & 80.0(5.6) \\ \\[0.10ex]
& \multirow{2}*{MF2}
 &1&$13-32$ & 21.0/16 & 0.47659(65) & 0.3873(8) & 0.01169(16) & 3.59(54) \\
&&2&$8-23$  & 14.1/12 & 0.5699(57)  & 0.4977(65)& 0.0361(16)  & 100.4(6.9)\\ \\[0.10ex]
& \multirow{2}*{MF3}
 &1&$13-18$ & 0.9/2 & 0.5289(28) & 0.4051(33) &  0.0152(7)  & 5.7(2.3) \\
&&2&$7-17$  & 6.7/7 & 0.6058(50) & 0.5013(62) &  0.0370(15) & 119.8(6.8) \\ \\[0.10ex]
& \multirow{2}*{MF4}
 &1&$13-32$ & 19.5/16 & 0.5727(46) & 0.4169(62) &  0.0176(13)  & 14.1(6.8) \\
&&2&$7-17$  & $8.6/7$ & 0.6399(73) & 0.5052(92) &  0.0380(23)  & 116.0(10.3) \\
\hline\hline
\end{longtable*}

\subsection{Finite-size effects}
\label{SubSec: Effect of finite lattice spacing}

We employ the following relations:
\begin{eqnarray}
\label{eq:Disp_Two_Cont_k}
\sqrt{s}  &=& \sqrt{ E_{MF}^2 - P^{\star2} }, \cr
p^{\star2}  &=& \frac{s}{4} - m_\pi^2  ,
\end{eqnarray}
in the Lorentz transformation for the invariant mass $\sqrt{s}$,
the energy of $\pi\pi$ system in the moving frame $E_{MF}$
and the center-of-mass scattering momentum $p^\star$.
Equation~(\ref{eq:Disp_Two_Cont_k}) is only suitable up to the truncation errors.
Rummukainen and Gottlieb suggest~\cite{Rummukainen:1995vs}
\begin{eqnarray}
\cosh( \sqrt{s} ) &=&
\cosh(E_{MF}) - 2\sin^2\left(\frac{P^\star}{2} \right), \cr
2\sin^2 \left(\frac{p^\star}{2}\right)    &=& \cosh\left( \frac{\sqrt{s}}{2} \right) -\cosh(m_\pi)
\label{eq:Disp_Two_Lat_k}
\end{eqnarray}
to reduce this truncation error.
Recently, we extended them to a two-particle system with arbitrary masses~\cite{Fu:2012gf}.

We discern no obvious difference of the ultimate  results
due to the selection of the energy momentum relations
(\ref{eq:Disp_Two_Cont_k}) or (\ref{eq:Disp_Two_Lat_k}) within the statistics,
especially for lattice ensemble with a smaller lattice space $a$ or large spatial extent $L$.
For these reasons, we computed the $\sqrt{s}$ and $p^\star$ by the continuum
relation~(\ref{eq:Disp_Two_Cont_k}).

\section{Analysis}
\label{Sec:analysis}
We are now in a position to use the scattering phases $\delta_1$
to secure the $\rho$ resonance parameters.
Moreover, since we have six sets of lattice data at hand,
we can follow the pioneering work of the ETMC Collaboration~\cite{Feng:2010es}
to discuss the pion mass dependence on $\rho$ resonance parameters.
After chiral extrapolation to the physical point,
the desired physical quantities can be obtained.

\subsection{Resonant parametrizations}
To estimate the two-pion energies, we use the well-known effective
range formula~\cite{Agashe:2014kda}
\begin{equation}
\label{eq:effective_range_formula}
\tan{\delta_1}=\frac{g^2_{\rho \pi\pi}}{6\pi}
\frac{p^3}{ \sqrt{s}(M_R^2- s)}  ,  \quad
p           = \sqrt{ \frac{s}{4} - m_\pi^2 } .
\end{equation}
where the Mandelstam variable $s$ is denoted by the center-of-mass energy of the $\pi\pi$ system $E_{CM}$ through $s = E_{CM}^2$.
This enables a fit for two unknown quantities: the coupling constant $g_{\rho \pi\pi}$
and the resonance mass $M_R$ from the lattice-determined $p$-wave scattering phase $\delta_1$.
The $\rho$ decay width is then calculated through
\begin{equation}
\label{eq:decay_width}
\Gamma_{\rho} = \Gamma_R(s)\Bigg|_{s=M_R^2} =
\frac{g^2_{{\rho}\pi\pi}}{6\pi}\frac{\left(\frac{M_R^2}{4} - m_\pi^2\right)^{3/2}}{M_R^2} .
\end{equation}
Equations~(\ref{eq:effective_range_formula}) and (\ref{eq:decay_width}) offer us a way
to acquire $\rho$ range parameters by examining the dependence of $\delta_1$ on $\sqrt{s}$.

\begin{figure*}[htp!]
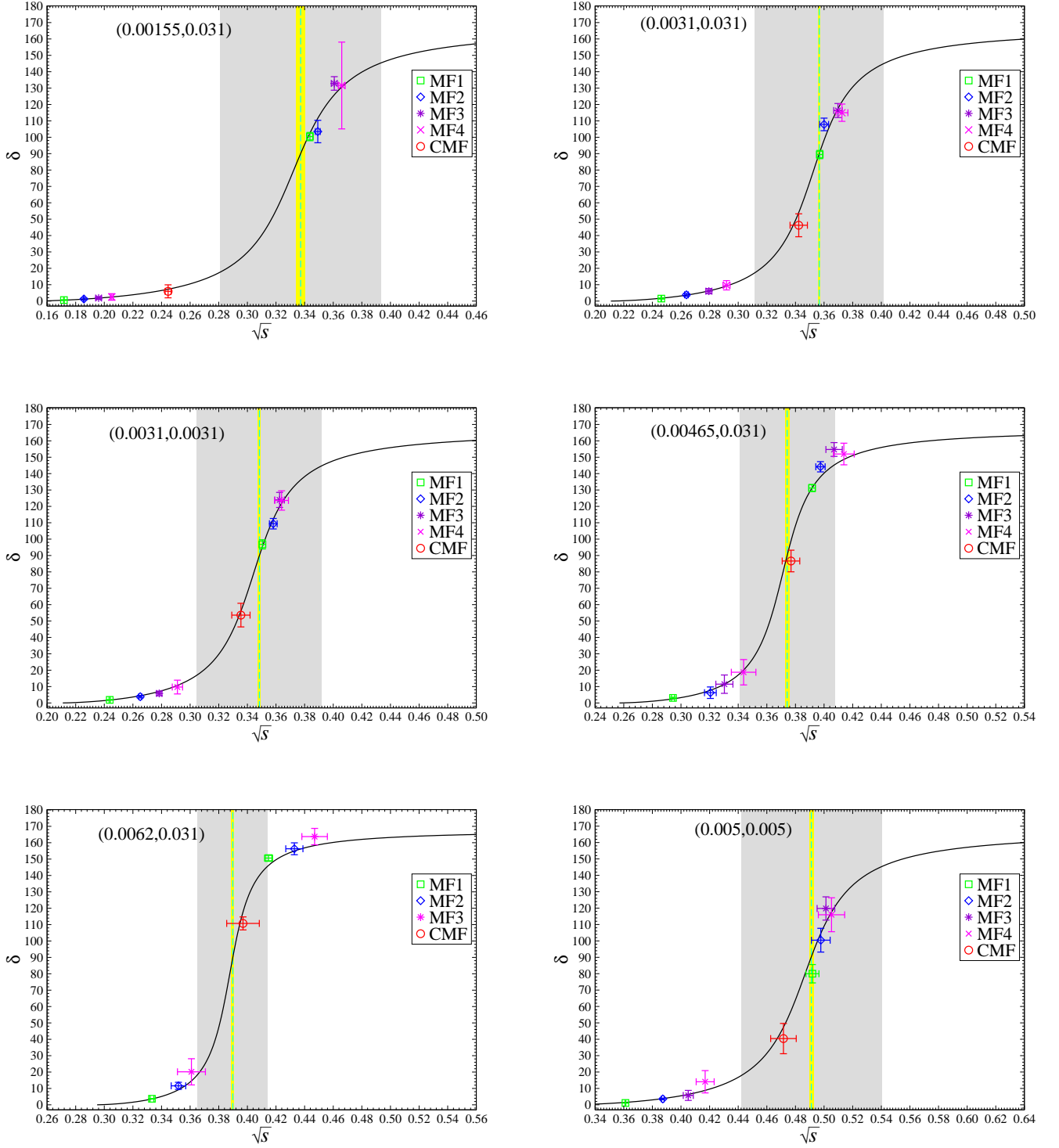

\begin{center}
\includegraphics[width=8.2cm]{00155_As.eps}\hspace{1cm}
\vspace{1.0cm}
\includegraphics[width=8.2cm]{0031_As.eps}
\vspace{1.0cm}
\includegraphics[width=8.2cm]{7405_As.eps}\hspace{1.0cm}
\includegraphics[width=8.2cm]{465_As.eps}
\includegraphics[width=8.2cm]{0062_As.eps}\hspace{1cm}
\includegraphics[width=8.2cm]{715_As.eps}
\end{center}
\caption{\label{fig:scattering_phase_A1}
Results of the scattering phase shifts and effective range formula fits
for the ensembles $(0.0031,0.031)$ (upper left), $(0.0031,0.0031)$ (upper right),
$(0.00465,0.0031)$ (middle left), $(0.0062,0.031)$ (middle right),
$(0.005,0.005)$ (bottom left),  and $(0.005,0.05)$ (bottom right).
The scattering phase shifts are calculated in the CMF, MF1, MF2, MF3 and MF4, respectively.
The solid black curves exhibit the central values of the effective range formula fits.
The dashed cyan lines display the resonance masses $am_\rho$,
the narrow yellow bands display their uncertainties,
and the resonance regions $am_\rho \pm a\Gamma$ are shown in the shadowed grey boxes.
}
\end{figure*}

\subsection{Extraction of the resonance parameters}
\label{SubSec:Scattering Phase Shift and Decay Width}

For six MILC lattice ensembles, we obtained seven or nine separate energy levels,
and we can then extract seven or nine $p$-wave scattering phases $\delta_1$
from the relevant invariant mass $\sqrt{s}$; these are shown in Fig.~\ref{fig:scattering_phase_A1}.
To extract the resonance mass $m_\rho$ and the coupling constant $g_{\rho\pi\pi}$
from a single lattice ensemble,
the seven or nine $p$-wave scattering phases $\delta_1$ are then fitted
with the effective range formula denoted in Eq.~(\ref{eq:effective_range_formula}).\footnote{
Other parametrizations have been recently discussed
for the $\rho$ resonance in Refs.~\cite{Dudek:2012xn,VonHippel:1972fg,Pelaez:2004vs,Li:1994ys}.
Additionally, the RQCD Collaboration found that all the resonant
masses obtained from other parametrizations  are in perfect agreement with those from Breit-Wigner~\cite{Bali:2015gji}.
}
The corresponding fits for six lattice ensembles are also exhibited in Fig.~\ref{fig:scattering_phase_A1}.
The fitted  $m_\rho$ in $\rm MeV$ and $g_{\rho\pi\pi}$
are summarized in Table~\ref{tab:rho_mass_width},
where the statistical errors of the lattice spacing $a$ are also added in quadrature.

Once the fitted values of the $g_{\rho\pi\pi}$ and $m_\rho$ in lattice units are acquired,
the decay width $\Gamma_\rho$ in lattice units can be estimated by Eq.~(\ref{eq:decay_width}),
where the uncertainties are solely estimated
from the statistical errors of both $g_{\rho\pi\pi}$ and $am_\rho$.
The calculated $\Gamma_\rho$ is also listed in Table~\ref{tab:rho_mass_width},
where the statistical errors of the lattice spacing $a$ are also added in quadrature.
It is worth mentioning that $\Gamma_\rho$ is mainly determined by the $\pi\pi$-phase space;
consequently, this number derived from the different quark masses turns out to be different.\footnote{
It is interesting and important to note that for the $(0.0031,0.031)$ and $(0.0031,0.0031)$ lattice ensembles,
the pion masses are almost the same and $L$ is identical;
nonetheless, the strange sea quarks for two lattice ensembles are quite different.
However, the discrepancies of the resonance mass $m_\rho$ for two ensembles are clearly noticed,
which indicates the influence of the strange sea quark.
}
Note that the lattice-calculated $\Gamma_\rho$ for the larger quark masses are
significantly smaller than the experimental value quoted
by the Particle Data Group (PDG) $\Gamma_{\rho}^{\rm phys} = 147.8(9)$~\cite{Agashe:2014kda}.
To make our demonstrations of these results more intuitive,
the resonance masses and the resonance regions are offered graphically  in Fig.~\ref{fig:scattering_phase_A1}.

\begin{table}[htp!]
\caption{\label{tab:rho_mass_width}
Summary of the fitted $\rho$-meson mass $m_\rho$ and the effective
coupling constant $g_{\rho\pi\pi}$ for the six MILC lattice ensembles.
The relevant estimated decay widths $\Gamma_\rho$ are also listed,
where the statistical errors of the lattice spacing $a$ are also considered.
The last block shows fit quality $\chi^2/{\rm DOF}$.
}
\begin{ruledtabular}
\begin{tabular}{|c|c|c|c|c|}
$\rm Ensemble$ & $m_\rho$ (MeV) & $g_{\rho\pi\pi}$ & $\Gamma_\rho$ (MeV)  & $\chi^2/\mathrm{DOF}$\\
\hline
$(0.00155,0.031)$  & $791(10)$ & $5.82(35)$ & $128(16)$   & $3.73/7$   \\
\hline
$(0.0031,0.0031)$  & $827(8)$  & $6.09(21)$ & $104(7)$    & $3.55/7$   \\
\hline
$(0.0031,0.031)$   & $836(8)$  & $6.03(28)$ & $106(10)$   & $9.09/7$    \\
\hline
$(0.00465,0.031)$  & $875(9)$  & $5.88(21)$ & $76.4(5.7)$ & $12.3/7$  \\
\hline
$(0.0062,0.031)$   & $915(9)$  & $5.80(17)$ & $57.6(3.6)$ & $12.5/5$ \\
\hline
$(0.005,0.005)$    & $840(8)$  & $5.90(19)$ & $90.8(5.7)$ & $2.16/7$ \\
\end{tabular}
\end{ruledtabular}
\end{table}

\newpage

\subsection{Comparison with other results}
\begin{figure}[htp!]
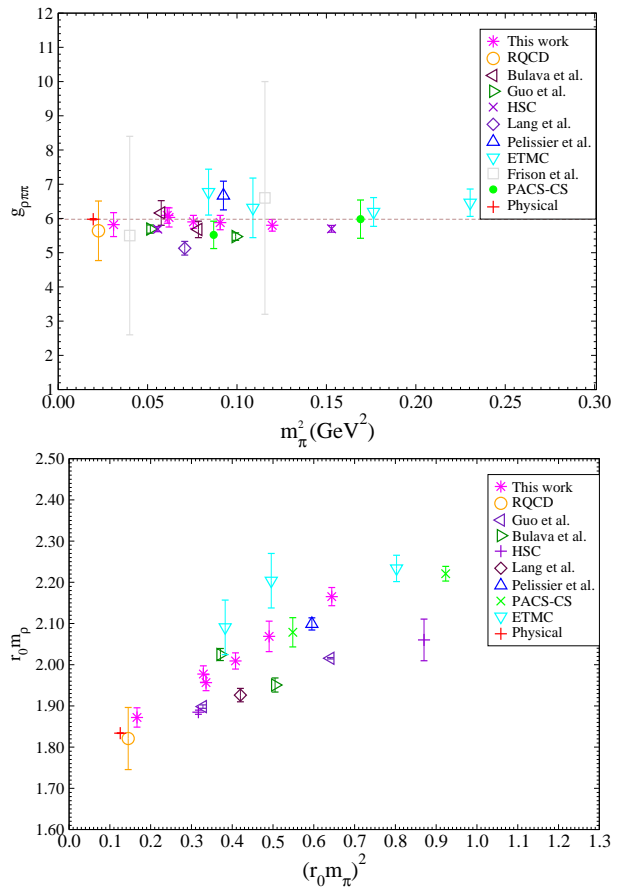

\begin{center}
\includegraphics[width=8.0cm,clip]{grpp.eps}
\includegraphics[width=8.0cm,clip]{rrho.eps}
\end{center}
\caption{\label{fig:comp}
Comparison of our results with other lattice studies.
The top panel shows the effective coupling constant $g_{\rho\pi\pi}$
as a function of $m_\pi^2$.
The bottom panel shows the resonance mass $m_\rho$,
where $m_\rho$ and $m_\pi$ are scaled with Sommer scale $r_0$~\cite{Sommer:1993ce}.
The red pluses indicate the corresponding PDG values.
}
\end{figure}

In Fig.~\ref{fig:comp}, we compare our lattice results of $\rho$ resonance parameters
from the MILC Asqtad-improved staggered fermions ($2+1$ or $3$ flavors)
with some other lattice studies:
the improved Wilson fermions ($2$ flavors, CP-PACS~\cite{Aoki:2007rd}),
the maximally twisted mass fermions ($2$ flavors, ETMC~\cite{Feng:2010es}),
the tree-level improved clover-Wilson
fermions ($2$ flavors, Lang {\it et al.}~\cite{Lang:2011mn}),
the nHYP-smeared clover fermions ($2$ flavors, Pelissier and Alexandru~\cite{Pelissier:2012pi}),
the nonperturbatively $O(a)$-improved Wilson fermion ($2$ flavors, PACS-CS~\cite{Aoki:2011yj}),
the anisotropic Clover Wilson fermions ($2+1$ flavors, HSC~\cite{Dudek:2012xn,Wilson:2015dqa}),
the improved Wilson fermions ($2+1$ flavors, Frison {\it et al.}~\cite{Frison:2010ws}),
the anisotropic Wilson clover fermions ($2+1$ flavors, Bulava {\it et al.}~\cite{Bulava:2015qjz}),
the nHYP-smeared clover fermions ($2$ flavors, Guo {\it et al.}~\cite{Guo:2016zos}),
and the nonperturbatively improved Wilson fermions ($2$ flavors, RQCD~\cite{Bali:2015gji}).
The top panel of Fig.~\ref{fig:comp} plots the effective coupling constant $g_{\rho\pi\pi}$
and the bottom panel shows the resonance mass $m_\rho$.
The systematic uncertainty for the determination of the lattice spacing
is added to the statistical error in quadrature.
It is important to note that our lattice results obtained with staggered fermions
are reasonably consistent with those using other actions, which have quite different systematics.

The effective coupling constant $g_{\rho\pi\pi}$ is dimensionless,
and thus practically has a weak quark mass dependence.
We also observe that the results of $g_{\rho\pi\pi}$ from all the lattice studies
are almost consistent in top panel of Fig.~\ref{fig:comp}.
Our results of $g_{\rho\pi\pi}$ are well consistent
with other lattice studies and were determined with the similar precision.
Indeed, the stability of the results for  $g_{\rho\pi\pi}$  with respect to other magnitudes was anticipated in Ref.~\cite{Chen:2012rp}.

It is worth mentioning that the resonance mass $m_\rho$ is very sensitive to the pion masses.
In order to avoid the artificial systematic error from
the determination of the lattice spacings,
which are used to measure $m_\rho$ and $m_\pi$ in lattice units,
it is proper to adopt dimensionless quantities to
compare the resonance mass $m_\rho$ with each other, and it is natural
to use the Sommer scale $r_0$~\cite{Sommer:1993ce}.
The bottom panel of Fig.~\ref{fig:comp} shows the resonance mass $m_\rho$,
where $m_\rho$ and $m_\pi$ are both scaled with the Sommer scale $r_0$~\cite{Sommer:1993ce}.
The lattice spacings $a$ and  $r_0$ for six lattice ensembles used in the present work
have been professionally determined by MILC in Refs.~\cite{Bernard:2010fr,Aubin:2004wf,Bazavov:2009bb},
we can directly quote these results;\footnote{
This work benefits a lot from the MILC Collaboration; without their published data,
we could not launch this work.
}
Lang {\it et al.} determined the lattice spacing by inputting the $r_0=0.48$~fm~\cite{Lang:2011mn} and
Pelissier and Alexandru fixed the lattice spacing
by setting $r_0=0.5$~fm~\cite{Pelissier:2012pi}.
The value of $r_0$ for the ETMC gauge configuration was determined  to be $r_0/a = 5.32(5)$~\cite{Baron:2009wt},
the PACS-CS gauge configuration has been reported as $r_0/a = 5.427(51)$~\cite{Aoki:2008sm}
and that for HSC as $r_0= 0.454$~fm~\cite{Lin:2008pr}.
The RQCD determined the lattice spacing by setting  $r_0 = 0.501$~fm~\cite{Bali:2015gji,Bali:2012qs},
and Guo {\it et al.} usually set $r_0 = 0.5$~fm~\cite{Guo:2016zos}.
We should remark at this point that the relevant PDG value in Fig.~\ref{fig:comp}  is just
scaled with MILC's determinations of $r_0$ on the same lattices of this work~\cite{Bernard:2010fr,Aubin:2004wf,Bazavov:2009bb}
since it is reasonably compatible with Sommer's continuum extrapolation of $r_0$ for the published $N_f > 2$ determinations~\cite{Sommer:2014mea}.

Nonetheless, from the bottom panel of Fig.~\ref{fig:comp},
large differences for the resonance mass $m_\rho$ are still discerned.
Note that there has been no attempt so far with a continuum limit extrapolation.
As pointed out in Ref.~\cite{Aoki:2011yj},
there exist some other possible issues to interpret this discrepancy,
such as the discretization error,
the influence of the strange sea quark (as we already discern in Sec.~\ref{SubSec:Scattering Phase Shift and Decay Width}),
the issue of the isospin breaking and the reliability of the effective range parametrization, ad son forth.
In any case, the robust extraction of the resonance mass $m_\rho$
definitely need more lattice simulations in the vicinity of the physical point,
as well as a continuum limit extrapolation.


\subsection{Quark mass dependence}
So far, only the ETMC Collaboration has discussed the quark mass dependence of
$\rho$ resonance parameters~\cite{Feng:2010es};
the other lattice works have been studied with one pion mass or two pion masses~\cite{Aoki:2007rd,Aoki:2011yj,Gockeler:2008kc,Feng:2010es,Frison:2010ws,Pelissier:2012pi,Dudek:2012xn,Lang:2011mn,Bulava:2015qjz,Guo:2016zos,Wilson:2015dqa,Bali:2015gji}.
Since, six quark masses are used in the present study,
we are now in a position to examine the pion mass
dependence of the $\rho$ resonance parameters.

The quark mass dependence of the $\rho$ resonance parameters are discussed
with effective field theory in Ref.~\cite{Jenkins:1995vb}.
The pion mass dependence of the $\rho$ resonance mass $m_\rho$ and
$\rho$ decay width $\Gamma_\rho$ can be generally expressed as~\cite{Jenkins:1995vb}\footnote{
Note that $m_\rho$ and $\Gamma_\rho$ are statistically correlated,
indicating that the coefficients $C_{m_i}$ and $C_{\Gamma_i}$ ($i=1,2$)
are not independent from each other.
Therefore, following the strategy of Refs.~\cite{Djukanovic:2009zn},
Xu {\it et al} introduced the complex pole parameter $Z$ to fit their data~\cite{Feng:2010es}.
}
\begin{eqnarray}
\label{eq:fit_M3_mass}
m_\rho      &=& m_\rho^0+C_{m1} M_\pi^2+C_{m2}m_\pi^3+O(m_\pi^4)\;,\\
\vspace{0.6cm}
\Gamma_\rho &=& \Gamma_\rho^0+C_{\Gamma1}m_\pi^2+C_{\Gamma2}m_\pi^3+O(m_\pi^4)\;.
\label{eq:fit_M3_gamma}
\end{eqnarray}

\begin{figure}[ht]
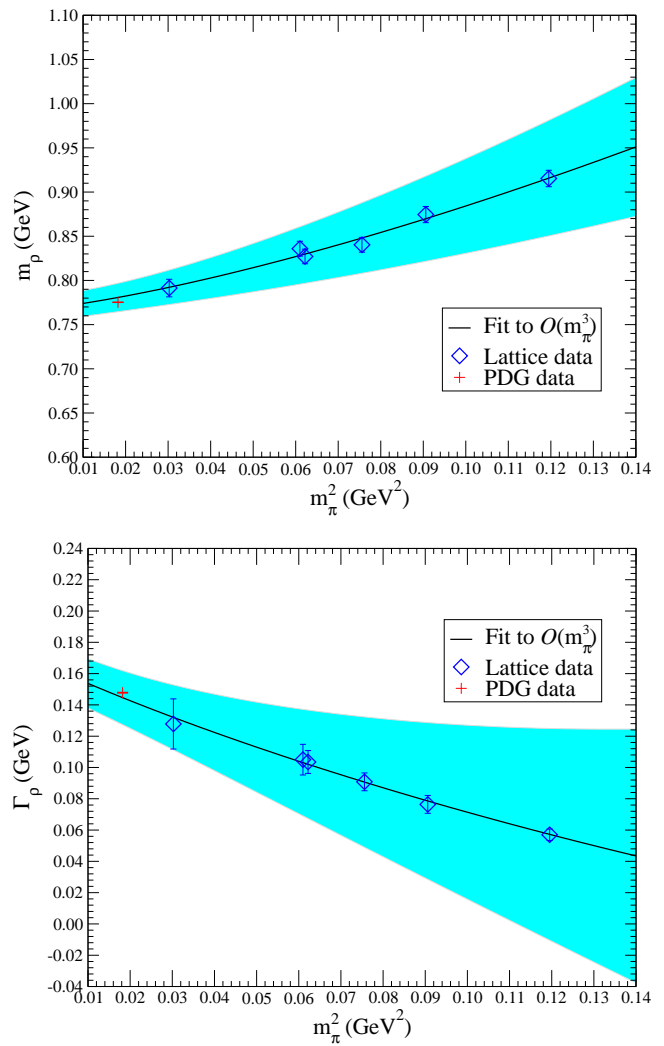

\begin{center}
\includegraphics[width=8.5cm]{mrho_mpi.eps}\vspace{0.3cm}
\includegraphics[width=8.5cm]{grho_mpi.eps}
\end{center}
\caption{\label{fig:M_R}
The lattice-measured $\rho$ resonance parameters as the functions of the pion mass squared.
The upper panel exhibits the $\rho$-meson resonance mass and
the lower panel shows the $\rho$-meson decay width.
The cyan bands correspond to the fits to our six data points using the
Eq.~(\ref{eq:fit_M3_mass}) and Eq.~(\ref{eq:fit_M3_gamma}), respectively,
and the solid black curves are the central values of the corresponding fits.
The red plus points indicate the relevant PDG values.
}
\end{figure}

In the top panel of Fig.~\ref{fig:M_R}, we display the $\rho$ resonance mass as a function of $m_\pi^2$,
together with a fit to Eq.~(\ref{eq:fit_M3_mass}); these are also summarized on the left side of Table~\ref{tab:fitting_mass_width}.
After the chiral extrapolation to the physical point,
we obtain the physical $\rho$ resonance mass $m_{\rho;\rm phys} = 780(16)$~MeV,
where the uncertainty is solely estimated by the fitted statistical errors of the three coefficients in Eq.~(\ref{eq:fit_M3_mass})
that are listed in the left side of Table~\ref{tab:fitting_mass_width}.
It is obvious that our physical $\rho$ resonance mass $m_{\rho;\rm phys}$
is in good agreement with the PDG value of the $\rho$-meson mass
$m_{\rho;\rm PDG} = 775.26(25)$~MeV~\cite{Agashe:2014kda} within the statistical errors,
which is indicated by the red plus point in the top panel of Fig.~\ref{fig:M_R}.

As explained in Ref.~\cite{Feng:2010es},
since the ETMC Collaboration carried out the lattice calculations
at the relatively large pion masses (from $290$ to $480$~MeV),
their obtained physical $\rho$ resonance mass $m_{\rho;\rm phys}$
is relatively high compared to the PDG value even using the $O(q^4)$ extrapolations.
On the other hand, this work carries out a study with the relatively small pion mass
(from $176$ to $346$~MeV), and with the more lattice ensembles.
Note that the RQCD Collaboration recently worked at nearly physical quark masses~\cite{Bali:2015gji}.

\begin{table}[htp!]
\caption{\label{tab:fitting_mass_width}
Summary of the $\rho$ resonance mass $m_\rho$ and $\rho$ decay width $\Gamma_\rho$
fitting with Eqs.~(\ref{eq:fit_M3_mass}) and ~(\ref{eq:fit_M3_gamma}), respectively.
The values of $ m_\rho^0$ and $\Gamma_\rho^0$ are given in units of GeV,
those of $C_{m1}$  and $C_{\Gamma1}$ are given in units of $\rm GeV^{-1}$,
and those of $C_{m2}$  and  $C_{\Gamma2}$ are given in units of $\rm GeV^{-2}$.
The corresponding fits result in fit qualities $\chi^2/{\rm DOF}$=$2.06/3$, $0.33/3$, respectively.
}
\begin{ruledtabular}
\begin{tabular}{|cc|cc|}
Fit of $m_\rho$ to  &  Eq.~(\ref{eq:fit_M3_mass})  & Fit of $\Gamma_\rho$ to  & Eq.~(\ref{eq:fit_M3_gamma})    \\
\hline
$ m_\rho^0$  & $0.768(14)$ & $\Gamma_\rho^0$ & $0.166(15)$   \\
\hline
$C_{m1}$     & $0.38(42)$  & $C_{\Gamma1}$   & $-1.33(44)$    \\
\hline
$C_{m2}$     & $2.49(96)$  & $C_{\Gamma2}$   & $1.21(96)$    \\
\end{tabular}
\end{ruledtabular}
\end{table}

In practice, the decay width $\Gamma_\rho$ can be estimated through Eq.~(\ref{eq:decay_width}),
where the statistical errors are estimated from the statistical errors of
both $g_{\rho\pi\pi}$ and $m_\rho$.
Therefore, our lattice-extracted decay widths $\Gamma_\rho$ definitely indicate a union of the two factors.
In the bottom panel of Fig.~\ref{fig:M_R}, we exhibit the decay width $\Gamma_\rho$
as a function of the pion mass squared,
along with a fit to Eq.~(\ref{eq:fit_M3_gamma}); these are also summarized in the right side of Table~\ref{tab:fitting_mass_width}.
Since Eq.~(\ref{eq:decay_width}) naturally regresses
to $\Gamma_\rho = m_\rho g^2_{\rho\pi\pi} / (48\pi)$ in the chiral limit,
it often leads to a good value of $\Gamma_\rho$ with the better value of $m_\rho$.
Moreover, the error of $g_{\rho\pi\pi}$ will be more quickly propagated
in the $\Gamma_\rho$ than that of $m_\rho$.
After the chiral extrapolation to the physical point,
our physical $\rho$ decay width $\Gamma_{\rho;\rm phys} = 144.6(17.3)$~MeV,
where the uncertainty is solely estimated by the fitted statistical errors of the three coefficients in Eq.~(\ref{eq:fit_M3_gamma})
that are listed in the right side of Table~\ref{tab:fitting_mass_width}.
Our physical $\rho$ decay width $\Gamma_{\rho;\rm phys}$,
is slightly lower than the PDG value $\Gamma_{\rho;\rm PDG} = 147.8(0.9)$~MeV~\cite{Agashe:2014kda},
but it is in reasonable agreement with the PDG value within the statistical errors,
which is indicated by the red plus point in the bottom panel of Fig.~\ref{fig:M_R}.
Nonetheless, it is worth mentioning that our lattice-measured
$\rho$ resonance parameters are obviously much less accurate
than the PDG values~\cite{Agashe:2014kda}.

\section{Conclusions and outlook}
\label{Sec:Conclusions}
In this work, we for the first time employ the $N_f=2+1$ or $3$ flavors of the MILC Asqtad-improved staggered fermions at pion masses ranging from $176$ to $346$~MeV
to carry out the lattice computation of the $p$-wave $I = 1$ $\pi\pi$ scattering phase shifts
near the $\rho$  resonance region.
At all the pion masses, the physical
kinematics for the $\rho$-meson decay, $m_\pi/m_\rho<0.5$, is satisfied.
Additionally, from Table~\ref{tab:MILC_configs},
we note that our lattice volumes all have $m_\pi L > 4$;
thus, finite-size effects are negligible,
and the L\"uscher formulas are perfectly satisfied,
since the finite-size effects are exponentially suppressed with the combination $m_\pi L$.
In particular, we marked out the resonance region
by simultaneously adopting five Lorentz frames (CMF, MF1, MF2, MF3, and MF4).

Moreover, we for the first time investigated $\rho$ resonance parameters
with the moving-wall source technique~\cite{Kuramashi:1993ka,Fukugita:1994ve}, a nonstochastic source method.
We have shown that the lattice computation of
the $p$-wave scattering phase for the $I=1$ $\pi\pi$ system
using the moving-wall source and then the estimation of
the decay width of the $\rho$ meson are feasible and effective,
 andcan be comparable with the stochastic source method~\cite{Aoki:2007rd,Aoki:2011yj,Gockeler:2008kc,Feng:2010es,Frison:2010ws,Pelissier:2012pi,Dudek:2012xn,Wilson:2015dqa,Bali:2015gji},
 or its variants (the distillation method, etc.~\cite{Lang:2011mn,Bulava:2015qjz,Guo:2016zos}).
Most of all, we extracted the $\rho$-meson decay width
from the scattering phase data and demonstrated that
it is reasonably comparable with the $\rho$-meson decay width
from PDG within the statistical errors.

We evaluated the scattering phase from the seven or nine energy levels
for the six lattice ensembles by the L\"uscher finite-size methods.
The scattering phases are fitted with the effective range formula
to extract the $\rho$ resonance mass $m_\rho$, the decay
width $\Gamma_\rho$ and the effective coupling $g_{\rho\pi\pi}$.
Despite not considering the inherent relation
between $m_\rho$ and $\Gamma_\rho$, we conducted a fit
guided by the effective field theory to our results at six pion masses.
This provided an alternative means of the chiral extrapolation to the physical point.

After the chiral extrapolation to the physical point,
we obtain the physical $\rho$-meson mass $m_{\rho,\mathrm{phys}}$=$780(16)$~MeV,
which is in agreement with the experimental value $m_{\rho;\rm PDG} = 775.26(25)$~MeV~\cite{Agashe:2014kda},
and the decay width $\Gamma_{\rho,\mathrm{phys}}$=$144.6(17.3)$~MeV,
which is slightly low relative to the experimental value
$\Gamma_{\rho;\rm PDG} = 147.8(0.9)$~MeV~\cite{Agashe:2014kda}.
Moreover, our results are compatible with most recent lattice studies~\cite{Aoki:2007rd,Aoki:2011yj,Gockeler:2008kc,Feng:2010es,Frison:2010ws,Pelissier:2012pi,Dudek:2012xn,Lang:2011mn,Bulava:2015qjz,Guo:2016zos,Wilson:2015dqa,Bali:2015gji}.
It is obvious that our lattice computations cannot yet
match the experimental accuracy.

With the development of better algorithms, more efficient codes,
and an increase in computational resources,
the lattice calculations of the $\rho$ resonance parameters
with large $L$, small pion, and fine lattice will become possible,
which will make the lattice simulation more accurate~\cite{Bali:2015gji} (see the Appendix for more details).
With this aim in mind, our ongoing lattice studies
will be carried out with the MILC superfine gauge configuration
($a\approx 0.6$~fm, $L=48$, and beyond),
 and even with the MILC ultrafine gauge configuration
($a\approx 0.45$~fm and $L=64$).
These studies will include several lattice spacings, which enables us to make a continuum limit extrapolation.
Furthermore, working close to the physical pion mass with large $L$ or very fine lattice
are crucial for lattice investigations of the scattering processes involving thresholds,
e.g., X(3872), $D\bar{D^*}$, and beyond~\cite{Lang:2014yfa}.

\section*{Acknowledgments}
This work is partially supported by both the National Magnetic Confinement Fusion Program
of China (Grant No. 2013GB109000)
and the Fundamental Research Funds for the Central Universities (Grant No. 2010SCU23002).
We would like to express our deep appreciation to the MILC Collaboration for
allowing us to use the MILC gauge configurations and MILC codes.
We would like to thank the NERSC for providing a convenient platform to download the MILC gauge configurations.
We sincerely thank Carleton DeTar for his introduction to Karl Jansen's lattice works about six years ago
and indoctrinating me in the necessary theoretical knowledge
and computational skills for this work.
We especially thank Eulogio Oset for his enlightening and constructive comments and corrections.
The author must express my respect to Geng Liseng, Liu Chuan, and Chen ying
for reading this manuscript and give some useful comments.
Once again, we cordially express our boundless gratitude to prof. Hou qing's strong support,
prof. He Yan's contributing to us two powerful workstations,
and prof. Fujun Gou's vigorous supports,
otherwise, it is impossible for us to conduct this expensive work,
and have an opportunity to fulfill it.
We should show our appreciations to profs. He Yan, Huang Ling, Wang Jun, An Zhu, Liu Ning and Fu Zhe,  etc. for donating us
enough removable hard drives to store the quark propagators for this work.
We also express gratitude to the Institute of Nuclear Science and Technology, Sichuan University,
and Chengdu Jiaxiang Foreign Language School,
from which the computer resources and electricity costs for this study were furnished.
Numerical calculations for this work were carried out at
both PowerLeader Clusters and AMAX, CENTOS, HP, ThinkServer workstations.

\appendix

\section{The noise-to-signal ratio of the correlator}
\label{appe:sn}
In Ref.~\cite{Fukugita:1994ve}, the noise-to-signal ratio of
the two-point function at zero momentum evaluated with $N_{cfg}$
independent gauge configurations is estimated as
\begin{equation}
R_{SN}^2 \propto \sqrt{\frac{1}{N_{\rm cfg}L^3}} \exp[{(m_M-m_\pi)t}]
\label{eq:Rsn1}
\end{equation}
where $L$ is the lattice spatial dimension, and $m_M$ is the desired meson mass.
The superscript in $R$ indicates that this is the {\it two}-point function.

It is straightforward to extend this expression to
the two-point function at nonzero momentum ${\mathbf p}$
\begin{equation}
R_{SN}^2 \propto \sqrt{\frac{1}{N_{\rm cfg}L^3}}\exp[{(E_M-m_\pi)t}]
\label{eq:Rsn2}
\end{equation}
where the meson energy $E_M=\sqrt{m_M^2+{\mathbf p}^2}$,
${\mathbf p}=\tfrac{2\pi}{L}{\mathbf n}$.
In fact, this expression can be inferred from the analytical arguments in Refs.~\cite{Parisi:1983ae,DellaMorte:2012xc}.
In practice, in order to improve the statistics, the correlators
are calculated from a given number of time slices ($N_{\rm slice}$).
The corresponding noise-to-signal ratio can be roughly evaluated as
\begin{equation}
R_{SN}^2 \propto \sqrt{\frac{1}{N_{\rm cfg}N_{\rm slice}L^3}}\exp[{(E_M-m_\pi)t}].
\label{eq:Rsn3}
\end{equation}
Here we crudely assume that the calculations from different time slices are independent.
In our concrete numerical calculations~\cite{Fu:2014rea}, we indeed adjust the
values of $N_{\rm slice}$ to obtain the relevant masses with the desired precision;
at the same time, we found $R_{SN}^2 \propto 1/(N_{\rm cfg})^{\alpha}$,
where the exponent $\alpha=0.4\sim0.5$.
Therefore, the relationship of the noise-to-signal ratio
with $N_{\rm slice}$ in Eq.~(\ref{eq:Rsn3}) is approximately satisfied.

The dramatic deterioration of the signal as the momentum increases
is shown in Fig.~2 of Ref.~\cite{DellaMorte:2012xc}.
This quite impressive result indicates that the expected asymptotic
behavior given in Eq.~(\ref{eq:Rsn3}) is generally met.
We should remark at this point that, in practice,
the asymptotic trend given in Eq.~(\ref{eq:Rsn3}) can effectively
guide us how to improve the relevant statistical errors.

For the $\pi\pi$ scattering (two pions with the momentum ${\mathbf p}$
and ${\mathbf q}$, respectively),
the noise-to-signal ratio of the four-point function can be straightforwardly generalized
as~\cite{Fukugita:1994ve}
\begin{equation}
R_{SN}^4 \propto \sqrt{\frac{1}{N_{\rm cfg}N_{\rm slice}L^3}}
\exp[{(E_\pi({\mathbf p})+E_\pi({\mathbf q}) - 2m_\pi)t}],
\label{eq:Rsn4}
\end{equation}
where the energy $E_\pi({\mathbf p}=\tfrac{2\pi}{L}{\mathbf n}) = \sqrt{m_\pi^2+\tfrac{4\pi^2}{L^2}{\mathbf n}^2}$
and $E_\pi({\mathbf q}=\tfrac{2\pi}{L}{\mathbf m}) = \sqrt{m_\pi^2+\tfrac{4\pi^2}{L^2}{\mathbf m}^2}$.

According to the above analytical discussions, we can readily deduce that the most efficient way
to improve the relevant noise-to-signal ratios is to use very fine gauge configurations
where the temporal lattice spacing $a_t$ and the spatial lattice spacing $a_s$ are pretty small ($a_s=a_t$),
since the energy and the mass are measured in lattice $a_t$ units $a_t m$ and $a_t E$; and equivalently,
the use of the anisotropic gauge configurations,
where $a_t$ is much smaller than $a_s$, is also a powerful approach to
improve the relevant noise-to-signal ratios~\cite{Dudek:2012xn,Wilson:2015dqa}.
In addition, if we use the lattice ensembles with larger lattice spatial dimensions $L$,
and sum the  correlators over all the  time slices (i.e., $N_{\rm slice}=T$,
where $T$ is the lattice temporal dimension), the signals of the correlators
should also be significantly improved.


\end{document}